\documentclass[12pt]{iopart}
\usepackage{graphics}
\usepackage{epsfig}
\usepackage{citesort}

\usepackage{a4wide}

\usepackage{latexsym}
\usepackage{epsfig}
\usepackage{amssymb}
\newcommand{\bea}{\begin{eqnarray}}
\newcommand{\eea}{\end{eqnarray}}
\newcommand{\be}{\begin{equation}}
\newcommand{\ee}{\end{equation}}

\newcommand{\ra}{\rightarrow}

\newcommand{\ph}{\phantom}
\newcommand{\an}{\xi}
\newcommand{\nabu}{{}^{(3)}\nabla}

\begin{document}

%\title{Dynamics of Accelerating Cosmologies with an Anisotropic Equation of State}

% or perhaps:
%\title{Accelerating Cosmologies with an Anisotropic Equation of State: Dynamics of Background and Perturbations}

% or perhaps:
\title{Anisotropic Dark Energy: Dynamics of Background and Perturbations}

%\date{\today}

\author{Tomi Koivisto}
\mailto{T.Koivisto@thphys.uni-heidelberg.de}
\address{Institute for Theoretical Physics, University of Heidelberg, 69120 Heidelberg,Germany}
\author{David F. Mota}
\mailto{D.Mota@thphys.uni-heidelberg.de}
\address{Institute for Theoretical Physics, University of Heidelberg, 69120 Heidelberg,Germany}

\begin{abstract}
We investigate cosmologies where the accelerated expansion of the Universe is driven by a
field with an anisotropic equation of state. We model such scenarios within the Bianchi I
framework, introducing two skewness parameters to quantify the deviation of pressure from isotropy. 
We study the dynamics of the background expansion in these models. 
A special case of anisotropic cosmological constant is analyzed in detail. 
The anisotropic expansion is then confronted with  the redshift and angular distribution of the supernovae type Ia. In 
addition, we investigate the effects on the cosmic microwave background (CMB) anisotropies for which the main signature appears 
to be a quadrupole contribution.
We find that the two skewness parameters can be very well constrained. 
Tightest bounds are imposed by the CMB quadrupole, but there are anisotropic models which avoid this bound completely. 
Within these bounds, the anisotropy can be beneficial as a potential explanation of various anomalous cosmological 
observations, especially in the CMB at the largest angles. 
We also consider the dynamics of linear perturbations in these models. 
The covariant approach is used to derive the general evolution equations for cosmological perturbations 
taking into account 
imperfect sources in an anisotropic background. The implications for the galaxy formation are then 
studied. These results might help to make contact between the observed anomalies in CMB
and large scale structure and fundamental theories exhibiting Lorentz violation. 
\end{abstract}

\maketitle
%\newpage

\tableofcontents

\newpage

\section{Introduction}

There is a remarkable amount of observational evidence that the large-scale structure of the Universe resembles nearly the 
simplest and most symmetric imaginable system. Hence, it is well approximated by the flat geometry version of the 
Friedmann-Lema\^{i}tre models which are both homogeneous and isotropic. The fluctuations in the cosmic microwave background (CMB) 
are a crucial probe of these properties, and therefore the recent detections of some unexpected features in the CMB temperature 
anisotropies have raised a lot of speculations about the need to reconsider some of the basic cosmological 
assumptions. A hemispherical asymmetry has been reported \cite{Eriksen:2003db}. The angular correlation spectrum 
seems to be lacking power at the largest scales \cite{Hinshaw:2006ia}. The alignment of the quadrupole and 
octupole (the so called Axis of Evil \cite{Land:2005ad}) could also be seem as an extra-ordinary and unlikely 
result of statistically isotropic perturbations, even without taking into account that these multipoles happen 
also to be aligned to some extent with the dipole and with the equinox. The Axis of Evil does not show any 
correlation with the lack of angular power \cite{Rakic:2007ve}. Given the {a \it posteriori} nature of these 
considerations, the cosmic variance (and the inevitable arbitrariness in any statistics), the statistical 
significance of all these anomalies is indeed a debatable issue \cite{Magueijo:2006we,Land:2006bn,Copi:2005ff,Hajian:2006ud}.  

Various possible cosmological effects have been proposed as possible explanations for those anomalous
features. For instance, to introduce a preferred axis, one has to generate some isotropy breaking. According
to whether this occurs at an early time or at late times, one may classify the models into those in which
an originally isotropic CMB fluctuation is distorted on its way to us \cite{Gordon:2005ai} and into those whose
statistical anisotropy is imprinted in the primordial fluctuations \cite{Armendariz-Picon:2005jh}. A
primordial imprint has been suggested to be left by isotropization happening during inflation
 \cite{Gumrukcuoglu:2006xj,Ackerman:2007nb}, parity violating couplings  \cite{Alexander:2006mt} vector field
gradients \cite{Armendariz-Picon:2007nr} or non-standard spinors \cite{Bohmer:2007ut}. On the other hand, the apparent 
alignments could also be caused by local effects in our neighborhood. Notice, however, that the Axis of Evil has been 
argued to probably persist after foreground removal \cite{deOliveira-Costa:2006zj}. Local voids have been though suggested 
as its possible origin \cite{Inoue:2006rd} (and quite interestingly such voids could explain the present days acceleration too 
\cite{Enqvist:2006cg}). 

In the case CMB is distorted during the late acceleration, the signatures of anisotropy would automatically be seen at the smallest 
multipoles of the CMB, since the perturbation wavelengths corresponding to these angles enter inside the horizon at the same epoch 
that the dark energy dominance begins. Therefore one would have one coincidence problem less. The apparent 
statistical anisotropy has been associated with dark energy in the form of vector 
fields \cite{Armendariz-Picon:2004pm,Boehmer:2007qa,Libanov:2007mq,us}, shear viscous fluids \cite{Koivisto:2005mm} and elastic solids \cite{Battye:2006mb}. See 
also \cite{o1,o2,o3,o4} for other investigations. In general, if dark energy
is something else than the cosmological constant $\Lambda$-term, one expects it to have anisotropic stresses at 
least at the perturbative level. This is also a generic prediction of modified
gravity theories \cite{val1,val2,Koivisto:2006ie,Koivisto:2005yc,clifton,skordis,amar}. Moreover, it 
could be used to distinguish among scalar field theories in which the possible nonminimal coupling occurs only at the perfect fluid matter 
sector \cite{doug1,doug2,Koivisto:2005nr,luca}. Such perturbative shear
stresses have in general been attempted to be constrained through parameters 
describing directly the difference of the metric variables \cite{Caldwell:2007cw} and the viscous properties of generalized fluids 
\cite{Ichiki:2007vn,Mota:2007sz,brook1}. Unfortunately, the perturbative anisotropic stress could escape detection, when its influences are restricted to so large 
scales that they are only seen through the amplitude of the first few
multipoles of the CMB. In spite of that, in many specific models the Jeans scale of the 
(effective) dark energy is small enough for dark energy to form small-wavelength perturbations which might be probed by e.g. weak lensing experiments  
\cite{amendola,uzan,bruck,nunes}. Moreover, a {\it statistical} anisotropy due to dark energy would leave specific signatures which are not described alone by the amplitude of 
the CMB angular power spectrum $\sim \langle a_{lm},a^*_{lm} \rangle$, where $a_{lm}$ are the coefficients of the spherical expansion of the anisotropy. 
Instead, correlations between different multipoles $l$ are predicted, and also $m$-dependent patterns (whereas 
statistical isotropy would predict, on average, the same value regardless of $m$). The theoretical prediction from statistical anisotropy to the CMB is 
thus the nondiagonality of the matrix $\langle a_{lm}, a^*_{lm} \rangle$.  

Here we continue a previous study which investigated the implications and
origins of an anisotropic acceleration \cite{Koivisto:2007bp}. There we studied exact 
anisotropic but homogeneous solutions of the cosmological equations. We now
extend such investigation and generalise to the case where inhomogeneities are present and introduce a covariant framework for the study. The Bianchi I 
model is sufficiently simple to allow semi-analytical calculations, and it captures the 
basic effect present also in more complicated anisotropic 
models by featuring their important common property, direction-dependent expansion rates.
Such metric is useful to describe  magnetic fields \cite{Barrow:1997sy} or their Yang-Mills generalizations \cite{Barrow:2005df}. As is well known, the 
anisotropy due to such fields tends to decay. Usually 
the constraints inferred on the Bianchi type I models have indeed been for cases in which the universe may initially be anisotropic but will then 
isotropize. Solutions which do not isotropize would require anisotropic matter sources, which in addition should dilute slower than radiation or dust. 
There has been no evidence that such cosmological sources would exist. However, as the universe is presently believed to be dominated by some source with 
unexplained negative pressure, a question we would like to ask is whether this pressure could also be anisotropic. This is of interest even without
the apparent hints of statistical isotropy at the CMB. 

After the study of Barrow on constraining anisotropic stresses in the late universe \cite{Barrow:1997sy}, we note that very recently there has been also 
interest in employing Bianchi I to study the CMB. Campanelli {\it et al}  \cite {Campanelli:2006vb} describe magnetic fields present at the last 
scattering epoch resulting in an ellipsoidal universe. Similar effects have been proposed due to moving dark energy   
\cite{BeltranJimenez:2007ai}. Rodrigues has suggested that an anisotropic cosmological constant \cite{Rodrigues:2007ny} could result from an infrared noncommutative 
property of the spacetime. Longo has claimed that handedness of spiral galaxies is anisotropic  \cite{Longo:2007gr} and that this, 
(together with the Axis of Evil) could be a result of large-scale magnetic fields \cite{Longo:2007pc}. If true, this would be a strong further motivation 
for our study since our model could unify magnetic and dark energy fields by describing Yang-Mills field or the mutually coupled system. In addition to these
approaches, the anisotropic expansion rates have been exploited in inflationary considerations \cite{Bohmer:2007ut,Gumrukcuoglu:2006xj,Ackerman:2007nb,barrow1,barrow2}. 
Then the appearance of anomalous features can be consistent, but does not necessitate,
the existence of anisotropic sources. Inflation can smooth out initial anisotropies and 
inhomogeneities, but if the universe did not undergo too many e-folds of
inflation \cite{burgess,easson}, some traces of such possible anisotropies could now be detectable at the 
largest scales. On the other hand, if one assumes isotropic initial conditions for the inflation, anisotropies could originate during inflation from a 
nonstandard energy source. This could be the same as dark energy.  

We attack the problem at three fronts: phenomenologically, theoretically and observationally. In section \ref{basics} we
present our phenomenological description of anisotropic sources in terms of the skewness parameters $\delta$ and
$\gamma$, and derive the basic equations describing the model as a dynamical
system. We then check the generic asymptotic evolution of the universe in some simple cases. This is a unifying description which can be used
for any particular model of inflation or dark energy. Especially we contemplate the possibility of generalizing the usual $\Lambda$ term in such a way that it would 
correspond to a constant vacuum energy that may exert anisotropic pressure. 
%We find that vector fields provide a wide variety of possibilities which are also 
%phenomenologically interesting: time-like and space-like, isotropic and anisotropic, minimally coupled and having couplings in the gravity sector or in the matter 
%sector. 
Possibilities of unifying inflation and dark energy arise therefore naturally. Section \ref{bounds} is devoted to the study of the implications of these models 
to cosmological observations. Explicit constraints are derived from the luminosity distance-redshift relationship of the supernovae of type Ia (SNIa) and the amplitude 
of the quadrupole anisotropy in the CMB. We find that the additional parameters we introduced, $\delta$ and $\gamma$, are well constrained. 
Finally, the last part of this article considers the inhomogeneities in the models considered here.
In Section \ref{covariant} we briefly introduce our notation for the covariant formalism and then 
present the general equations describing anisotropic cosmologies. Then we apply these general results into some interesting cases. 
In section \ref{structure} we study then large scale structure formation under various specific assumptions about the model. 
We conclude in section \ref{conclusions}.

%
%\section{Formalism and Field Equations}
%

%\section{PART I : Homogeneous Cosmology}
\section{Anisotropic Equation of State}

We set up our framework within the Bianchi type I case. Usually the Bianchi models isotropize \cite{Ellis:1998ct}. This is, in particular, true of the Bianchi I model 
with a perfect fluid content.  Models that do not isotropize, have not been studied to such extent (though, for instance Ref. \cite{Buniy:2005qm} presents 
exact solutions of inflationary Bianchi I cosmologies in the presence of various anisotropic sources).
There has been good reasons for that: firstly, the modern cosmological observations, most importantly the satellite measurements of the CMB have established that the 
universe at large scales is isotropic to about one part in $10^{5}$. Secondly, one does not usually expect growing anisotropies, on the contrary to
have them usually requires some imperfect and thus exotic matter sources. Thus it would seem that only a
highly contrived and fine-tuned cosmology could exhibit the non-isotropizing property and consistency with the data.
However, during the last decade, as mentioned in the introduction, evidence has accumulated against both of the presumptions above.
Therefore, it is interesting to look for fixed points, especially scaling solutions, in the more general set-up allowing for 
direction-dependent expansion rates. It becomes then possible to see which kind of possibilities in general
exist for anisotropic expansion histories, and whether they could occur naturally (i.e. without fine-tuning). We therefore 
perform a dynamical system analysis. Then, we will also consider the specific case of a generalized $\Lambda$ term. 

\label{basics}

\subsection{Parametrization}

Consider a general fluid flow in a curved space-time. Assume the 4-velocity tangent to the fluid flow lines is 
$u^a$. This is the central object in the formulations below. Everything that follows will be defined in relation 
to this average flow. An useful tensor is then the projection tensor orthogonal to $u^a$, which is given by 
\be \label{projector}
h_{ab}=g_{ab}+u_a u_b.
\ee
The time derivative of any tensor $T^{ab \dots}_{\phantom{ab \dots}cd \dots}$ is understood as the derivative along the vector $u^a$, as
\be \label{timed}
\dot{T}^{ab \dots}_{\phantom{ab \dots}cd \dots} \equiv T^{ab \dots}_{\phantom{ab \dots}cd \dots; e}u^e.
\ee
Note that in general this differs from the derivative with respect to cosmic time, and agrees only for scalar fields. 
We also define $\nabu$, the covariant derivative operator obtained by projecting
the four-dimensional one as
\be
\nabu_e T^{ab \dots}_{\phantom{ab \dots}cd \dots} =
h^{a}_{\phantom{a}a'}
h^{b}_{\phantom{b}b'} \dots
h_{c}^{\phantom{c}c'}
h_{d}^{\phantom{d}d'}
h_{e}^{\phantom{e}e'} \dots
T^{a'b' \dots}_{\phantom{a'b'\dots}c'd' \dots; e'}.
\ee
The first covariant derivative of the $u^a$ is decomposed as follows:
\be
u_{a;b} = \omega_{ab} + \sigma_{ab} + \frac{1}{3}\theta h_{a b}-a_a u_b.
\ee
One calls $a_a=\dot{u}_a$ the acceleration, $\theta \equiv u^a_{\phantom{a};a}$ the expansion scalar, $\omega_{ab} = \omega_{[ab]}$
the vorticity tensor and $\sigma_{ab}= \sigma_{(ab)}$ the shear tensor. Both the shear tensor and vorticity tensor are traceless and orthogonal
to $u^a$. It is useful to define a representative length scale $\ell$ by
\be
\frac{\dot{\ell}}{\ell} = \frac{1}{3}\theta.
\ee
This is a covariant generalization of the usual e-folding. % $e^x$, see Eq.(\ref{efolding}).
Furthermore, the stress-energy tensor of a general fluid in curved spacetime is decomposed as follows:
\be \label{set}
T_{ab} = \rho u_a u_b + p h_{ab} + 2q_{(a}u_{b)} + \pi_{ab},
\ee
where $p$ is the pressure, $\rho$ is the energy density, $q_a$ is the
energy flux and $\pi_{ab}$ the fluid viscosity. 
%This is nothing but Eq.(\ref{decomp}) written explicitly and covariantly.

In the following we consider the choice $u^a = \delta^0$, so that the fluid flow is given by the time-like unit vector.
We also restrict ourselves to homogeneous cosmologies, and thus set any spatial gradient to zero. Furthermore, we assume
that the vorticity vanishes. These assumptions, summarized as
\be \label{bi_req}
u^a = \delta^a_0, \quad \nabu_a f = 0, \quad \omega_{ab} = 0
\ee 
characterize the Bianchi I cosmology. If we would further set $\sigma_{ab}=\pi_{ab}=0$, we would recover the 
Friedmann-Lema\^{i}tre-Robertson-Walker (FLRW) universe. 

In the Bianchi I spacetime, one needs three parameters to characterize the pressure of a fluid in all directions. In addition to
the isotropic part which one gets as the trace part of Eq.(\ref{set}),   
\be 
p = \frac{1}{3}h^{ab}T_{ab}, 
\ee
two more parameters are needed to fully determine the degrees of freedom in the traceless and symmetric tensor 
\be \label{pi_ab}
pi_{ab} = \left(h^c_{(a}h^d_{b)}-\frac{1}{3}h^{cd}\right) T_{cd}.
\ee
In Bianchi spacetime one may choose diagonal coordinate system, as we'll discuss later. 
It is occasionally claimed that one may consider the axisymmetric case without loss generality. 
It might then seem that one always choose the anisotropy axis parallel to one 
of the coordinate axes and thus, by considering a coordinate where anisotropy along one direction vanishes and due to the tracelessness requirement must
in the remaining directions have an opposite sign and equal magnitude, describe the anisotropic pressure with full generality. However, rotation of the
anisotropy axis would not preserve the diagonality of the system. The $3x3$ traceless symmetric tensor in (\ref{pi_ab}) has 5 independent entries, but
3 of them may be fixed by choosing the spatial coordinate system. Thus the anisotropic pressure is characterized by two degrees of freedom. 
This resembles the anisotropic property of gravitational waves, which, at the level of linear perturbations are conventionally parametrized by the two modes of 
polarization with respect to the wavevector of perturbations. Since in the homogeneous background there is no obvious definition for such a wavevector,
we instead use the following contractions for a fully covariant characterization of our anisotropic degrees of freedom:   
\be \label{dt}
D \equiv \frac{1}{\rho^2}\pi_{ab}\pi^{ab}, \quad T = \frac{1}{\rho^3}\pi^a_b\pi^c_a\pi^b_c. 
\ee
Higher contractions would not be independent of these, but instead $\pi^a_b\pi^c_d\pi_a^b\pi_c^d = 2D^2$,
$\pi^a_b\pi^c_d\pi^e_a\pi^b_c\pi^d_e = \frac{5}{6}TD$, etc. In the following we employ a combination of the 
quantities in (\ref{dt}) as
\be
\delta = -\frac{1}{3\sqrt{2}}\frac{D+F}{\sqrt{F}}, \quad \gamma = \frac{\delta}{2} \pm \sqrt{D^2-\frac{3\delta^2}{4}},
\ee
where
\be
F^3 = D^3 - 12T^2 + 2\sqrt{6}\sqrt{6T^4-D^3T^2}. 
\ee
From these covariant expressions the use of this parametrization is far from obvious. Let us therefore make an excursion
to the preferred coordinate system picked up by the symmetries of this model. In the diagonal system, where $\pi=(\pi_x,\pi_y,\pi_z)$, 
we have now
\bea \label{params}
\delta & = & -\frac{1}{3}(\pi_x-\pi_y)/\rho_X, \\
\gamma & = & -\frac{1}{3}(2\pi_x+\pi_y)/\rho_X. \nonumber
\eea
The use of these variables is mainly motivated by their straightforward interpretation as generalized equations of state (see the Appendix)
and concordance with previous notations. The parameters $\delta$ and $\gamma$ may be interpreted simply as differences of the pressure along
the $x$ and the pressure along the $y$ and $z$ axes, respectively. The axisymmetric cases are now seen to correspond to 
the three possible cases that either $\delta=0$, $\gamma=0$, or $\delta=\gamma$. One might get rid of the ambiguity by relabeling the axes in such a way if the 
pressure along any two axes is equal, the third one is called $x$ (a general rotation is not of course allowed by the symmetries).  
In the similar way, we will parametrize the shear stress of the metric as:
\bea \label{shears}
R & = &  \frac{3}{\theta}(\sigma_x-\sigma_y), \\
S & = &  \frac{3}{\theta}(2\sigma_x+\sigma_y). \nonumber 
\eea
Again, this is just a change of variables which can be convenient for some purposes. The dimensionless variables $R$ and $S$ might 
be interpreted as the fractional difference of expansion rates between the $y$ and $x$ (for $R$) and between the $z$ and $x$ (for $S$).  
The notations (\ref{params}) and (\ref{shears}) have been introduced in Ref. \cite{Barrow:1997sy}. Again, the straightforward 
interpretation of the shears in the metric sector, $R$ and $S$, is confined to the preferred coordinate system, but these quantities can be defined fully
covariantly in a way completely analogous to coefficients of the shear in matter sector, $\delta$ and $\gamma$ in Eq.(\ref{dt}).
  
%There are two pressure degrees of freedom in addition to the isotropic pressure, which we have to parametrize in some way. 
%There is then no obvious parametrization which would be preferable, 
%since the decomposition into the two degrees of freedom is not covariant. 
%The square of the stress $\pi^2 = 3(\delta^2+\gamma^2-\delta\gamma)\rho^2$ is covariant. 

\subsection{The matter content}

As the contents of the Universe we consider a two-fluid system, where only one of the fluids is responsible
for possible nonstandard (anisotropic) properties.  We have then a perfect fluid with an energy-momentum tensor (we denote it with index
$m$) and an imperfect energy-momentum tensor (this we denote with index $X$) 
\be \label{perfect_emt1}
T_{ab} = \rho_m(u_a u_b + w_m h_{ab}) + \rho_X(u_a u_b + w_X h_{ab}) + \pi_{ab},
\ee
where $w_m$ is the equation of state parameter, $w_m=1/3$ for radiation and $w_m=0$ for dust (and something
in between for their mixture). We also consider an imperfect fluid, which we allow to
have the most general energy-momentum tensor compatible with the assumptions (\ref{bi_req}). 
The shear stress $\pi_{ab}$ is due to this component solely, and therefore we drop the $X$ from this quantity.

The continuity equations are then given by the divergence of the energy-momentum tensors.
We let the two components also interact, and thus allow nonzero divergence for the individual components.
\be \label{cont_1}
\dot{\rho}_m + (1+w_m)\theta\rho_m = Q\rho_m, 
\ee
\be \label{cont_2}
\dot{\rho}_X + (1+w_X)\theta\rho_X + \sigma_{ab}\pi^{ab} = -Q\rho_m. 
\ee
where $Q$ describes the coupling. If there are no interactions, one finds that the matter density scales as 
$\rho_m \sim (\ell^3)^{-1-w_m}$. For most of the study, we will simply neglect the coupling between the $X$ and the isotropic component. 
However, we keep it for the sake of generality, since we will later be interested in finding cosmological scaling solutions.
%(which, inthe isotropic case, are known to exist only in the presence of a coupling).
% \qquad \rho \sim (\ell)^{-1-w_X}$. 

The system is of course not closed until we have determined the properties of the imperfect stresses $\pi_{\mu\nu}$.
This is equivalent to specifying the anisotropy parameters $\delta$ and $\gamma$ in Eq.(\ref{params}). 
In the case of viscous fluids \cite{Saha:2007if}, these stresses are related to the expansion factors. A covariant form for the
viscosity generated in the fluid flow is 
\be \label{l-l}
\pi_{ab} = \varsigma\left(u_{a;c}h^c_{\phantom{c}b}
+u_{b;c}h^c_{\phantom{c}a}
- \frac{2}{3}u^c_{\phantom{c};c} h_{ab}\right)
+ \zeta u^c_{\phantom{c};c} h_{ab}.
\ee
Now the conservation equations $T^{ab}_{\phantom{ab};a}=0$ reduce to the Navier-Stokes equations in the non-relativistic 
limit. Here $\varsigma$ is the shear viscosity coefficient, and $\zeta$ represents bulk viscosity. Heuristically, they describe the resistance of fluid flow to 
external stresses. Such result in acceleration (derivatives of the velocity four-vector) which are then compensated by effect of $\pi_{ab}$. The bulk viscous 
part reacts to volume-changing stresses as it is proportional to the expansion scalar; in cosmology it gives an extra Hubble friction. The shear viscous part 
corresponds to the the symmetrized and traceless part of the acceleration, thus reacting to an anisotropic stress. Ultimately the coefficients 
$\zeta$ and $\varsigma$ should be determined from the microscopic kinetic theory. Calculating the tensor (\ref{l-l}) explicitly 
with the help of the covariant definitions of section (\ref{covariant}), one can write the equations of state as
\be 
w = w_X - [3\zeta +\frac{2}{3\rho}(R+S)\varsigma]\frac{\theta}{3\rho}, \quad
\delta = \frac{2\varsigma R\theta}{9\rho}, \quad \gamma = \frac{2\varsigma S\theta}{9\rho}.
\ee
One notes even with nonzero shear viscous coefficient $\varsigma$, one does not generate shear in the metric. 
In fact, with a positive coefficient $\varsigma$ (which is required by second law of thermodynamics), the viscous property of the fluid in fact 
tends to decrease the amount shear\footnote{This may be seen by plugging the $\delta$ and $\gamma$ in the evolution equations 
(\ref{s_eq},\ref{r_eq}).}. 
A bulk viscosity \cite{Saha:2004qt,Brevik:2001ed,Brevik:2005bj,Brevik:2006wa}
could also have interesting consequences in this framework; however we do not focus further on the 
Navier-Stokes type of stress in the present study.

There are also several examples of imperfect cosmological sources with constant equations of state. 
A magnetic field along the $z$-direction \cite{King:2006cy}
could be modeled by $w_X=1/3$, $\delta=0$, $\gamma=-2$. A string with a constant tension along
the $z$-direction \cite{Raj:2007ug} could be described by $w_X=1/3$, $\delta=0$, $\gamma=1$.
An anisotropic cosmological constant 
%(for which the anisotropy arises from the
%deformation of the Poisson structure between the canonical momenta in such a way that the dynamics remain invariant under
%rescaling of the scale factors \cite{Rodrigues:2007ny}) 
is described by $w_X=-1$. We mention these possibilities in Table \ref{tab0}. We will return to these cases later. 
Note that our description with constant skewness parameters unifies and generalizes all of the mentioned models. Vector fields 
will be considered in much more detail in elsewhere \cite{us}; we have found that such models can also feature constant equations of 
state $\delta$ and $\gamma$ in special cases, though they in general are dynamical, depending on the potentials and couplings \cite{us}. 
We have gathered some other examples of imperfect matter in Table \ref{tab0}.

\begin{center}
\begin{table}
\begin{tabular}{|c|c|c|c|}
\hline
Model & $\delta$  & $\gamma$ & $w_X$  \\
\hline
Isotropic fluid & $0$ & $0$ & $w_X$ \\
\hline
String  & $0$ & $1$ & $1/3$ \\
\hline
Domain wall & $1$ & $1$ & $2/3$ \\
\hline
Magnetic field & $0$ & $-2$ & $1/3$ \\
\hline
Viscous fluid & $\frac{2\varsigma R\theta}{9\rho}$ & $\frac{2\varsigma S\theta}{9\rho}$ & 
$w_X-\zeta\frac{\theta}{\rho}$ \\
\hline
$\Lambda$, case I & $\frac{h}{3}\theta(2S-R)$ & $\frac{h}{3}\theta(S-2R)$ & $-1 $ \\
\hline
$\Lambda$, case II & $\delta$ (free) & $\gamma$ (free) & $-1-\frac{1}{3}(3-2R+S)\delta - \frac{1}{3}(3+R-2S)\gamma$ \\
\hline
\end{tabular}
\caption{\label{tab0} Models with anisotropic equations of state.}
\end{table}
\end{center}

\subsection{Evolution equations}

The generalized Friedmann equation can be written as
\be \label{friedmann1}
\theta^2 = 3(\rho_m + \rho_X + \sigma^2). 
\ee
We also define the dimensionless density fractions
\be \label{fractions}
\Omega_m \equiv 24\pi G\frac{\rho_m}{\theta^2}, \quad U \equiv \frac{\rho_X}{\rho_m+\rho_X}.
\ee
We introduce $\Omega_m$ to quantify the amount of matter in analogy to the standard cosmology as
the percentage of matter contribution to the square of the average expansion rate. Note that there is
an ambiguity when referring to matter densities since, the generalized Friedmann Eq. (\ref{friedmann1}) 
implying $$\Omega_m=(1-U)[1-3\sigma^2/\theta^2],$$ $\Omega_m$ is different from the matter density 
divided by the total density (which is $1-U$).

Our dynamical variables are then the density fraction $U$, and the two shear anisotropies $R$ and
$S$. As the time variable we define $x=\log{\ell}$. In the following a prime denotes derivative with respect to that.
The evolution equations for these are then
\be \label{sys1}
  U' =
       U\left(U -1\right)
            \left[\pi^{ab}\sigma_{ab}  + 3\left(w_X - w_m\right) \right] - UQ,
\ee
\be \label{sys2}
R' = -\left(3 + \frac{\theta'}{\theta}\right)R + 9\delta(\frac{3\sigma^2}{\theta^2}-1)U,
\ee
\be \label{sys3}
S' = -\left(3 + \frac{\theta'}{\theta}\right)S + 9\gamma(\frac{3\sigma^2}{\theta^2}-1)U.
\ee
One notes that only with non-vanishing skewness parameters, shear could be generated
also in an initially isotropic situation. Also, if $\delta=\gamma=0$, even initially nonzero $R$ or $S$ would typically
decay. Although the coupling term $Q$ does not complicate this system much, but appears only in the evolution
equation for $U$, its presence of course can change the dynamics completely. As mentioned earlier, we consider mainly $Q=0$ case in this paper,
but keep the coupling in the equations for the sake of generality.
The two last equations can be written explicitly in terms of solely $R$, $S$, and $U$ as
\be \label{s_eq}
  S' =
       \frac{3}{2}
       \left(1 - \frac{3\sigma^2}{\theta^2} \right)
       \left\{ S\left[U\left(w_X-w_m\right)+w_m-1\right]-6\gamma U \right\}
\ee
\be \label{r_eq}
  R' =
       \frac{3}{2}
       \left(1 - \frac{3\sigma^2}{\theta^2} \right)
       \left\{ R\left[U\left(w_X-w_m\right)+w_m-1\right]-6\delta U \right\}
\ee
In the following we will refer to an
effective equation of state, which we define as
\be \label{weff}
w_{eff} \equiv -\frac{2}{3}\frac{\theta'}{\theta}-1,
\ee
if $\sigma^2=0$, the universe expands as if dominated by a fluid with $w_m=w_{eff}$; if $\sigma^2 \neq 0$, the average of the three 
effective equations of state in each direction is $w_{eff}$.

\section{The Background as a Dynamical System} 

\subsection{Fixed Points in the Axisymmetric Case}
\label{scaling}

In the Friedmann-Lema\^{i}tre universe it has been proven difficult to address the coincidence problem by finding a
model entering from a matter-dominated scaling solution to an accelerating scaling solution.
Allowing the presence of three expansion rates perhaps opens up the possibility of finding more
such scaling solutions which might eventually help to understand the coincidence problem. 
It is also possible to model inflation within this context (an interesting case then describes inflation
as an era of primordial isotropization to the solution $R=S=0$ from a more general initial state). 
%Near the singularity the Kasner solution (\ref{empty}) is known to be important.
Therefore we consider the asymptotic behaviours of the
universe. In the simplest anisotropic case we may assume: 1) axisymmetry, $S=\gamma=0$, 2) constant
equations of state and skewness parameter $\dot{w}_m=\dot{w}=\dot{\delta}=0$, and
to begin with, also 3) no coupling between the components $Q=0$. To end the analysis, we briefly review the consequences of relaxing the assumption 
3). The results are summarized in the Table (\ref{tab1}), and also discussed in some detail below. Please note that now $R$ can be considered as a 
shorthand notation for the expansion-normalized invariant shear 
$$R = 3\sqrt{3}\frac{\sigma}{\theta},$$ and similarly $\delta$ corresponds to the 
density-normalized invariant shear $$\delta = \frac{\pi}{\sqrt{3}\rho}.$$  

Our system has 5 fixed points. With the assumptions 1)-3), there are two fixed points corresponding to matter domination. One is the
\begin{itemize}
\item FLRW solution
\be \label{flrw_s}
R=0, \quad U = 0.
\ee
We assume $w_m<1$. Then the FLRW solution is a stable node when
$w+\delta > w_m$, otherwise a saddle point\footnote{In this section $w$ corresponds to the equation of state in the $x$- and $y$ directions. Thus $w_X=w+\delta$, if 
$w_X$ stands for the average of the principal pressures.}. 

\item An anisotropically expanding empty universe
\be \label{empty}
R=\pm 3, \quad U = 0,
\ee
is also a fixed point.
This is a saddle point in the $R=+3$ case when $\delta<w-w_m$, and in the $R=-3$ case when $\delta>(w_m-w)/3$ and otherwise unstable. 
This is the so called Kasner solution, where the metric may be described by
\be \label{kasner1}
ds^2_{Kasner} = -dt^2+t^{2K_a}dx^2 + t^{2K_b}dy^2 + t^{2K_c}dz^2,
\ee
where the Kasner exponents satisfy 
\be \label{kasner2}
1=K_a+K_b+K_c=K_a^2+K_b^2+K_c^2. 
\ee
In the present case, due to the axisymmetry, the exponents $K_a=K_c$ are of course equal, but we wrote the general form for later convenience.
This solution is relevant near the singularity, $t \rightarrow 0$, where 
the matter may be neglected, $\rho_m/H^2 \sim \rho/H^2 \approx 0$. This solution may be realized in two ways\footnote{In terms of the expansion rates introduced in the appendix, the mentioned two possibilities correspond to $\dot{a}/a = 
-1/(3t),\qquad \textrm{and}\qquad \dot{b}/b = \dot{c}/c = 2/(3t)$, and $\dot{a}/a = 1/t,\qquad \textrm{and} \qquad \dot{b}/b = \dot{c}/c = 0$.}. 
Firstly, one may have expansion in two and contraction in one direction, so that $K_a=-1/3$ and $K_b=K_c=2/3$ (or let the two directions change roles 
$a \leftrightarrow b$ to get the solution with $R=-3$). 
%This corresponds to one contracting and two expanding directions.
Secondly, one may have two static directions and one expanding direction, corresponding to $K_a=-1/3$ and $K_b=K_c=0$ (or again let the two directions 
change roles $a \leftrightarrow b$ to get the solution with $R=-3$). 
%This corresponds to expansion in one direction.

\item An anisotropically expanding universe dominated by dark energy,
\be \label{cigar}
R= \pm 3, \quad U = 1,
\ee
is also a fixed point of our system. The pancake solution with $R=+3$ is
a saddle point if $w> 1+\delta$ or $w < w_m+\delta$, and the
cigar solution with $R=-3$ is a saddle point if $w+3\delta>1$ or
$w+3\delta<w_m$.
Otherwise they are unstable. On the average, the equation of state is for a stiff fluid, $w_{eff}=1$.
\item The anisotropic fixed point
\be \label{domi_s}
R = \frac{6\delta }{w+\delta-1}, \quad U =1.
\ee
represents a balanced expansion in different directions. 
Assuming here for simplicity that $w_m=0$ and restricting to the case
$w<w_m$, this can be a stable node if either $\delta+w>1$ or $$1+6\delta
+ \sqrt{1+16w(w-1)} > 2w \land 2w + \sqrt{1+16w(w-1)} > 1 + 6\delta.$$
We have a saddle point if either $$3\delta+w<1 \land 2w+\sqrt{1+16w(w-1)}
\le 1+6\delta$$ or $$1+\delta>w \land 1+6\delta+\sqrt{1+16w(w-1)} \le 2w.$$

\item The scaling solution
\be \label{scal_s}
R = \frac{3\left(w+\delta-w_m\right)}{2\delta}, \quad
U =
\frac{\left(w+\delta-w_m\right)\left(w_m-1\right)}{3\delta^2-2\delta\left(w-w_m\right)-\left(w-w_m\right)^2}.
\ee
is finally the fifth fixed point of our system.
Considering again only cases when $w<w_m=0$, this is a saddle point if
either $$\delta>0 \land 2w + \sqrt{1+16w(w-1)} \ge 1+6\delta$$ or
$$\delta<0 \land 1+6\delta+\sqrt{1+16w(w-1)}\ge 2w.$$ General stability conditions for this point are more complicated.
Instead of trying to find feasible analytic expressions for them, we resort to an empirical study. 
\end{itemize}

\begin{figure}[ht]
\begin{center}
\includegraphics[width=0.45\textwidth]{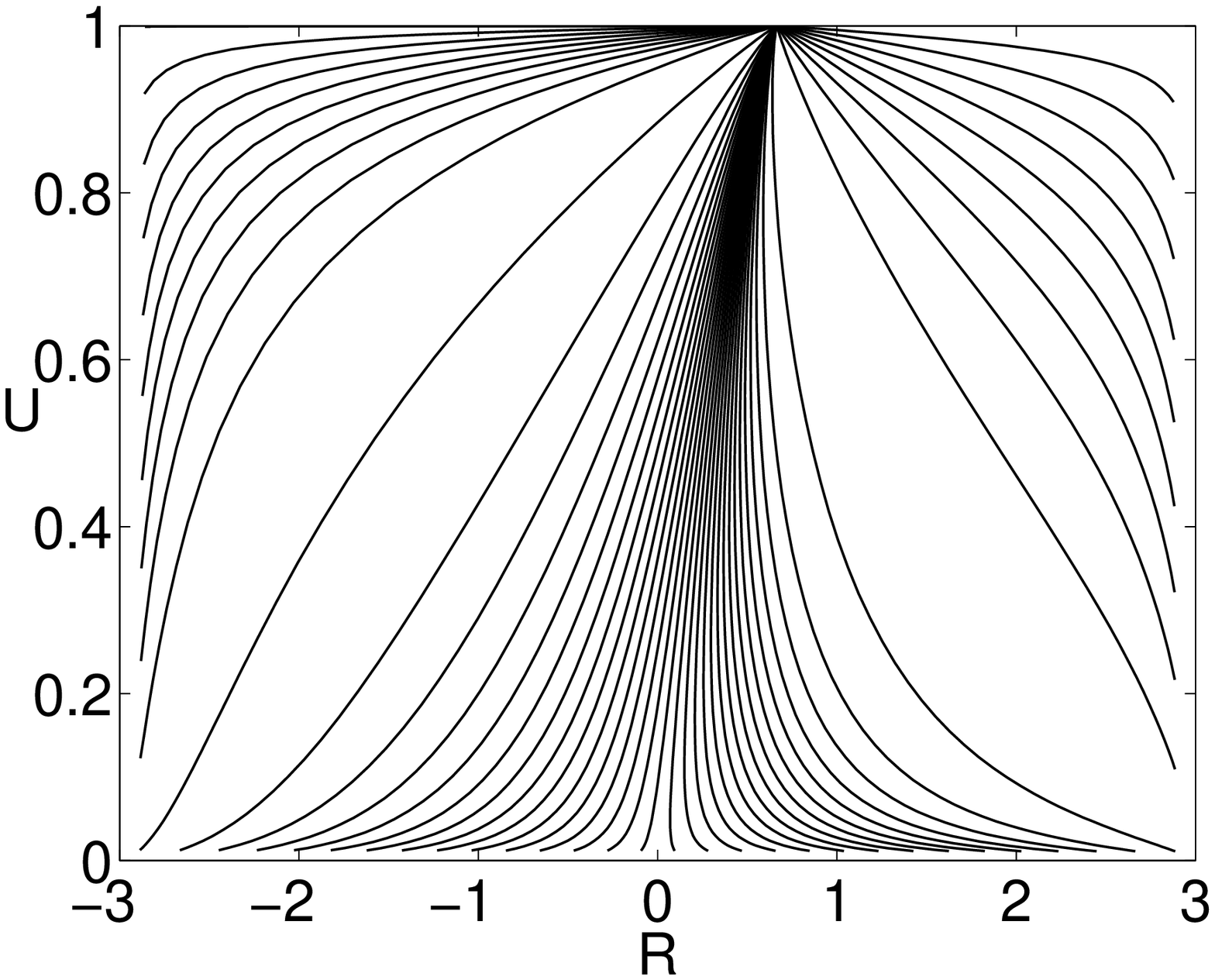}
\includegraphics[width=0.45\textwidth]{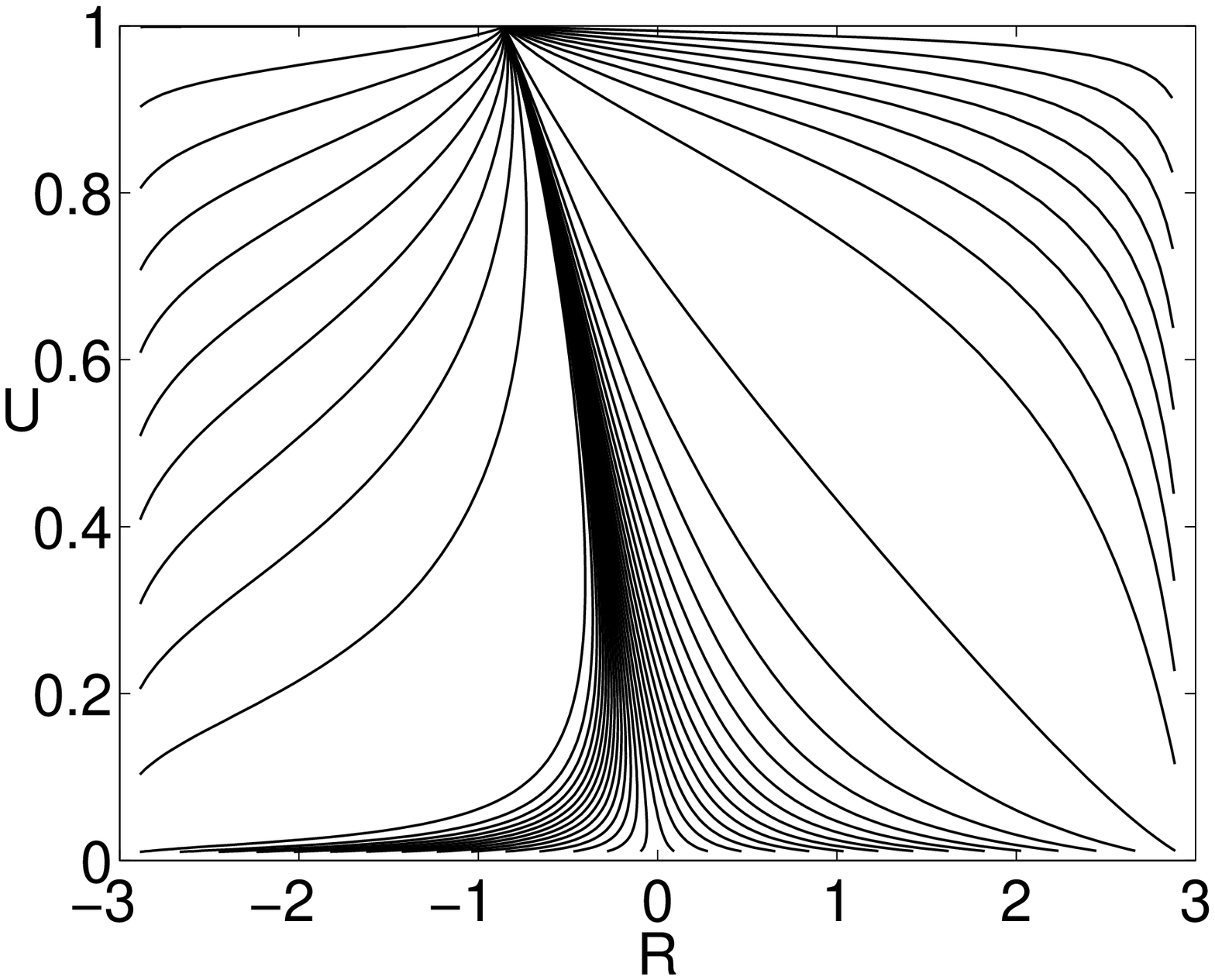}
\caption{\label{phases1} 
Phase portraits in the cases $\delta=-0.25$ (right) and $\delta=0.25$ (left) when $w=-1$.
The solution (\ref{flrw_s}) is at $R=U=0$, and the fixed points (\ref{empty}) and (\ref{cigar}) 
are at the four corners of the portraits.
%The Kasner solutions correspond to the vertical boundaries of he figure.
In both cases the fixed point (\ref{domi_s}) attracts trajectories from everywhere.}
\end{center}
\end{figure}

\begin{figure}[ht]
\begin{center}
\includegraphics[width=0.45\textwidth]{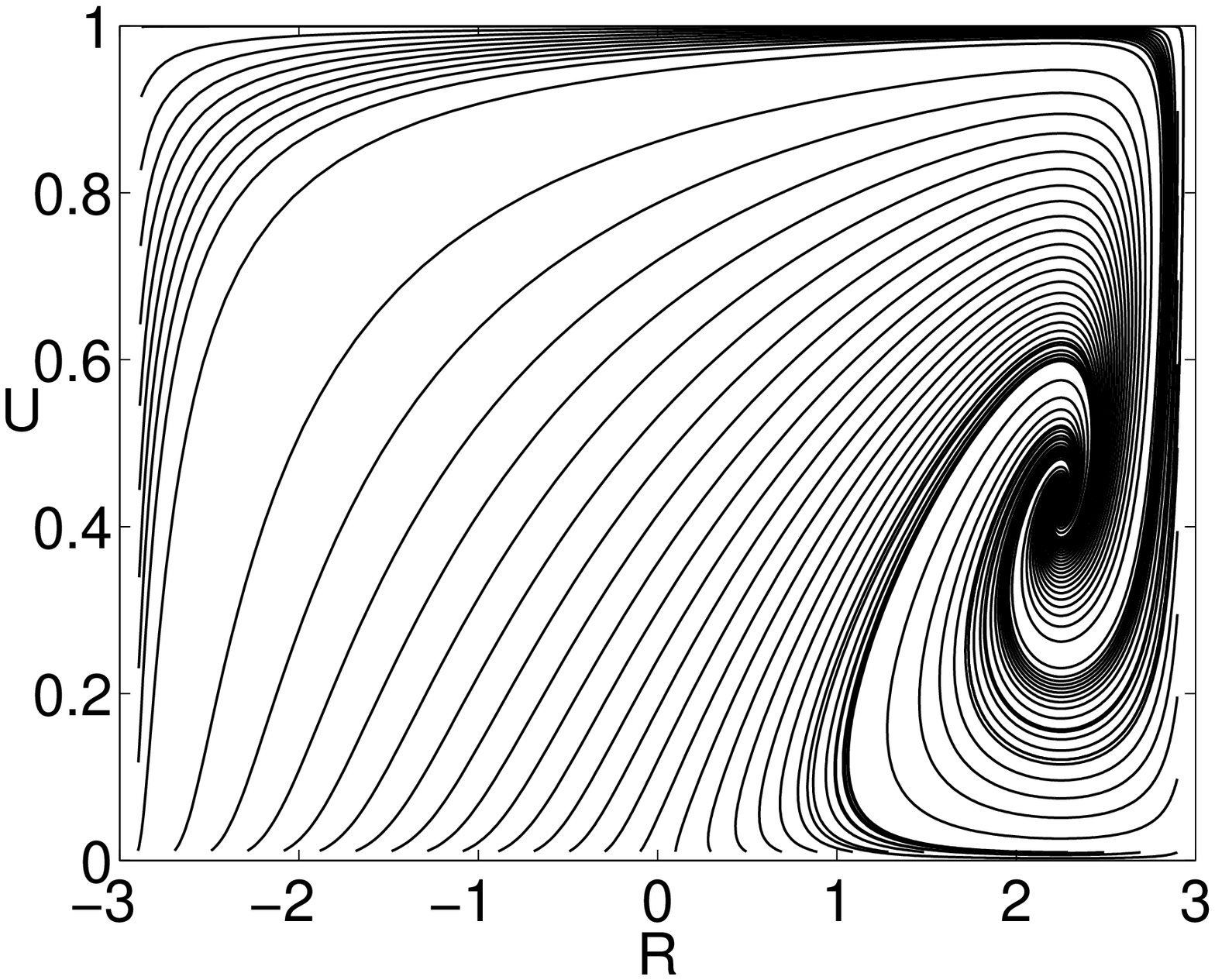}
\includegraphics[width=0.45\textwidth]{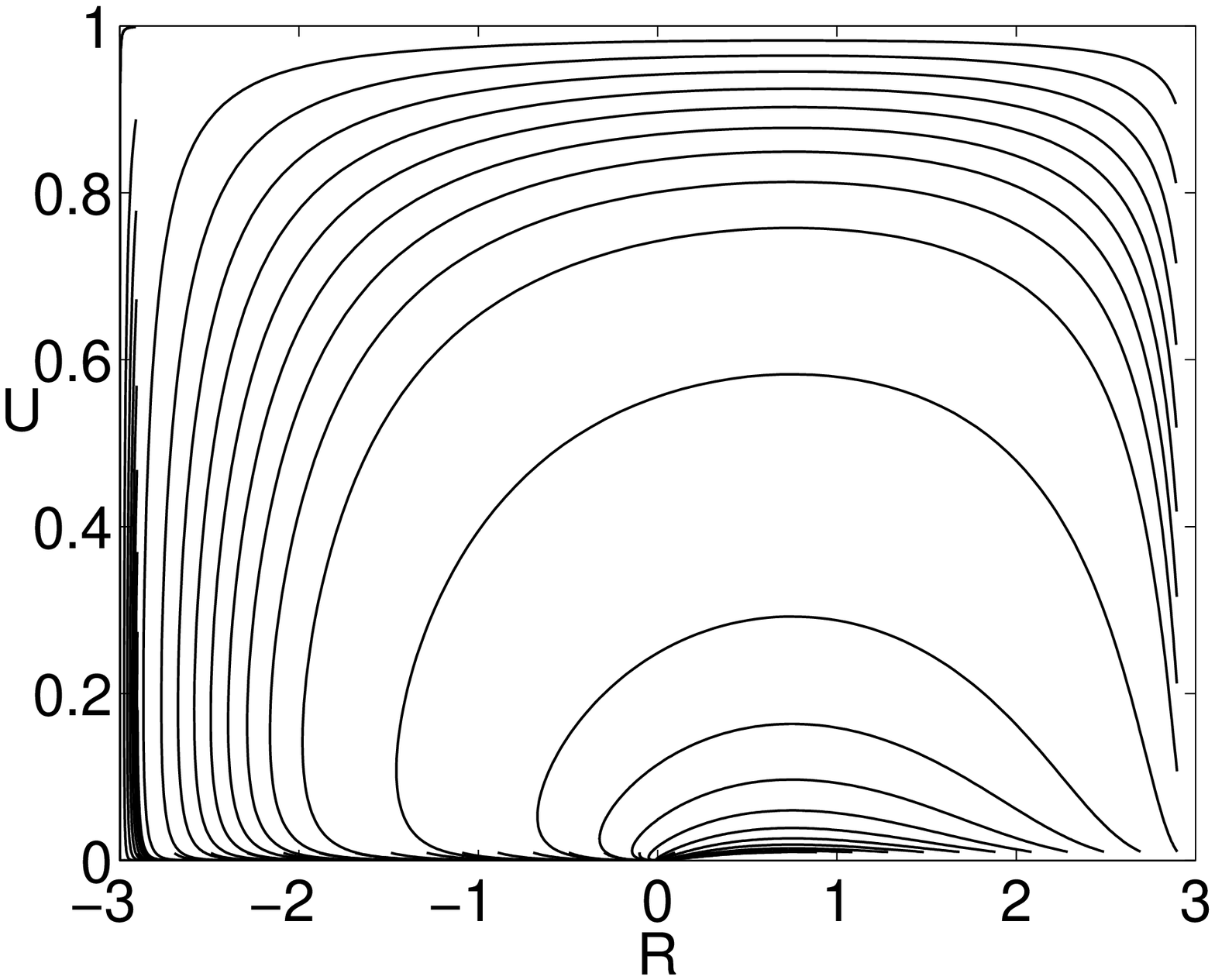}
\caption{\label{phases2} 
Phase portraits in the cases $\delta=-2$ (right) and $\delta=2$ (left) when $w=-1$.
The solution (\ref{flrw_s}) is at $R=U=0$, and the fixed points (\ref{empty}) and (\ref{cigar}) 
are at the four corners of the portraits.
%The Kasner solutions correspond to the vertical boundaries of the figure.
In right panel all trajectories lead to the scaling solution all (\ref{scal_s}). The left panel shows isotropizations
in the example with $w+\delta>w_m$.}
\end{center}
\end{figure}

\begin{figure}[ht]
\begin{center}
\includegraphics[width=0.7\textwidth]{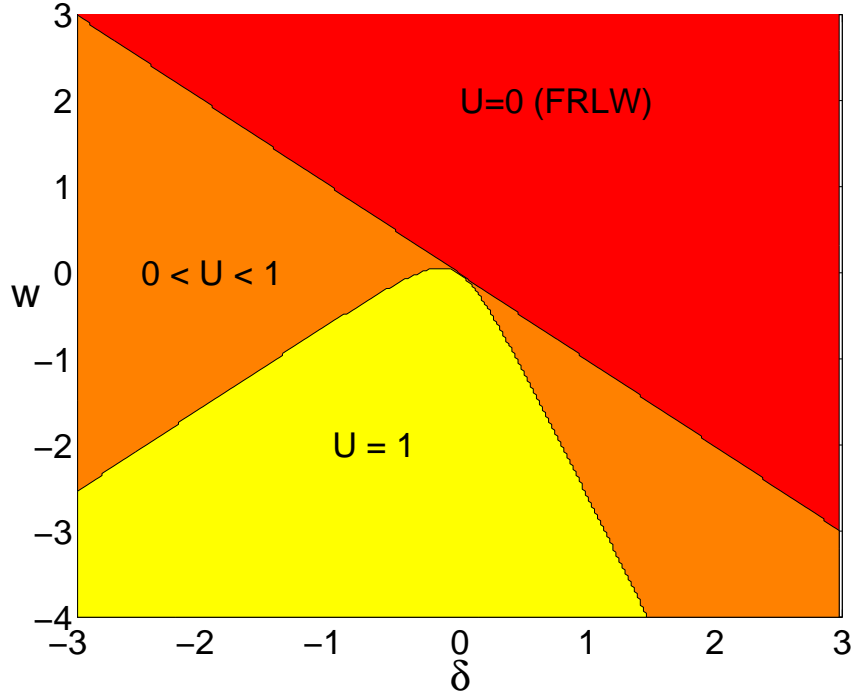}
\caption{\label{outcomes} The asymptotic state of the universe starting from an Einstein-deSitter stage.
The future fate depends on the dark energy properties $w$ and $\delta$ and is classified into three
possibilities. If $\delta+w < 0$, the isotropically expanding dust
domination (solution (\ref{flrw_s}) where $U=0$)
continues forever. Otherwise, the universe will end up expanding anisotropically
and either dominated by dark energy (solution (\ref{domi_s}) where $U=1$) or
exhibiting a scaling property (solution (\ref{scal_s}) where $0<U<1$). Fig. (\ref{phases1}) showed phase portraits
of two $U=1$ models, the right panel of Fig. (\ref{phases2}) showed an example of the $0<U<1$ case, and finally
the left panel of Fig. (\ref{phases2}) showed isotropizations in the $U=0$ (here we refer to the asymptotic value of $U$).}
\end{center}
\end{figure}

To study when the solutions are relevant for the dark energy problem, we use a dust-dominated FLRW universe,
$w_m=R=S=0$, as the initial condition. Then the outcome is uniquely determined by dark energy properties, $w$ and $\delta$. 
The numerical results, shown in Figs. (\ref{phases1},\ref{phases2}) and summarized in Fig. \ref{outcomes}, agree completely with our 
analytical considerations.

Let us finally consider relaxing the assumption 3) to include a coupling $Q$. We assume it is constant. This may be motivated by the analogy with scalar fields
(exponential interaction term in the Lagrangian results in constant $Q$ for scaling solutions), and can also be realized with vector fields having matter 
couplings, as we show in another publication with explicit examples \cite{us}.
Then the dark energy dominated anisotropic 
solution Eq.(\ref{cigar}) would turn into an anisotropic scaling solution
\be \label{cigar_g}
R = \pm 3, \quad U = 1 - \frac{Q}{3(w_m - w) + (3 \mp 6)\delta},
\ee
and
where the amount of matter is proportional to the strength of the coupling $Q$.

The scaling solution (\ref{scal_s}) generalizes in the presence of a coupling to
\bea \label{scal_g}
R = \frac{15\delta^2 + (-3 + Q + 3w)(w - w_m) + \delta\left(Q + 6w - 3(1 + w_m)\right) \pm \sqrt{r} }{4\delta(-1 +
\delta + w)},
\nonumber \\
U = \frac{9d^2 - \delta(3 + Q + 6w - 9w_m) -
     (3 + Q + 3w - 6w_m)(w - w_m) \pm \sqrt{r}}{6(3\delta + w - w_m)(\delta - w + w_m)}.
\eea
where
\bea
r = \left(-9\delta^2 + \delta (3 + Q + 6w - 9w_m) + (3 + Q + 3w - 6w_m)(w - w_m)\right)^2 - \nonumber \\
12(3\delta + Q + 3w - 3w_m)(3\delta + w - w_m)(w_m-1)(\delta - w + w_m).
\eea
Thus the coupling modifies both the relative amount of anisotropy and the relative amount of matter. An isotropic scaling 
solution is now possible with a nonzero coupling and $\delta=0$. We list a summary of the fixed points found in this 
subsection in the Table (\ref{tab1}).
\begin{center}
\begin{table}
\begin{tabular}{|c|c|c|c|c|c|}
\hline
Solution & Eqs      & $w_{eff}$ & $R$  & $U$ & Coupling \\
\hline
FLRW & (\ref{flrw_s}) & $w_m$      & $0$  & $0$ &  - \\
\hline
Empty & (\ref{empty}) & $1$      & $\pm 3$  & $0$ & - \\
\hline
DE 1 & (\ref{cigar}) & $ 1 \mp 2\delta $   & $\pm 3$  & $1$ & See Eq.(\ref{cigar_g}) \\
\hline
DE 2 & (\ref{domi_s}) & $1-\frac{4\delta^2}{\delta+w-1}$ & $\frac{6\delta }{w+\delta-1}$ & $1$ & - \\
\hline
Scaling & (\ref{scal_s}) & $w_m-\frac{4\delta^2(w_m-1)}{(3\delta + w - w_m)(\delta - w + w_m)}$ & 
$\frac{3\left(w+\delta-w_m\right)}{2\delta}$ & 
$\frac{\left(w+\delta-w_m\right)\left(w_m-1\right)}{3\delta^2-2\delta\left(w-w_m\right)-\left(w-w_m\right)^2}$ & 
See Eq.(\ref{scal_g}) \\
\hline
\end{tabular}
\caption{\label{tab1} Fixed points in the axisymmetric case, with constant equations of state $w$, $w_m$ and $\delta$. The 
effective equation of state $w_{eff}$ is defined in Eq.(\ref{weff}). The last column indicates whether the solution can be 
modified by a coupling. The fixed point corresponds to an anisotropic expansion unless $R=0$.}
\end{table}
\end{center}

The rest of the fixed points, Eqs.(\ref{flrw_s}), (\ref{empty}) and (\ref{scal_s}) would retain their form in the presence of a 
coupling term.

\subsection{An anisotropic Cosmological Constant}
\label{alambda}

We will show that there does not exist an anisotropic generalization of the cosmological constant with constant equations of state. Then we consider
two different classes of time-dependent cosmological terms with constant energy density. 
 
%For this purpose we focus briefly again to the background dynamics with our new covariant intuition.
The continuity equation (\ref{cont2}) for a component with general $w$, $\delta$ and $\gamma$ reads
\bea \label{continuity}
\dot{\rho}_\Lambda   + \theta\left(1+w_\Lambda\right)\rho_\Lambda + \pi_{ab}\sigma^{ab} & = & 0 \\
\dot{\rho}_\Lambda   + \theta\left[\left(1+w_\Lambda\right) + \frac{1}{3}\gamma\left(R - 2S\right) + \frac{1}{3}\delta\left(-2R+S\right) \right]\rho_\Lambda & = 
& 0.
\eea
We have added the subscript to emphasize we are discussing a cosmological constant term. In the second line we wrote open the component contributions to 
the invariant shear. As we then want $\dot{\rho}_\Lambda=0$, the square brackets term should vanish identically. If we have time-independent equations of state, we 
cannot consistently model an anisotropic $\Lambda$ term. Obviously, the constant terms in the square brackets should then vanish, and we should set $w_X=-1$.
Since there also appears $R$ and $S$, which in general depend on time, the linear combination
\be
X \equiv \frac{3}{\theta}\pi_{ab}\sigma^{ab} =  (R-2S)\gamma + (-2R+S)\delta
\ee
should however be zero. To see how this is impossible, one derives an evolution equation for $X$:
\bea \label{x_eq}
\dot{X}
& = & -\left(\theta+\frac{\dot{\theta}}{\theta}\right)X + \frac{6\pi^2}{\rho_X\theta} \\
& = & -\left(\theta+\frac{\dot{\theta}}{\theta}\right)X + 3\left[\gamma(\delta-2\gamma) + \delta(\gamma-2\delta)\right](\frac{3\sigma^2}{\theta^2}-1)\theta U.
\eea
The second term will always enforce $X$ to evolve unless $\delta=\gamma=0$. Note that the second term does not vanish since Eqs.(\ref{s_eq},\ref{r_eq}) 
set the shears evolving. 

As a summary, we note that in terms of our covariant notation, the two conditions mentioned above correspond to the two statements
\begin{itemize}
\item $w_\Lambda=1$, i.e. the isotropic part of the pressure should be equal to minus the energy
  density, $p_\Lambda=-\rho_\Lambda$. 
\item $X=0$, i.e. the anisotropic part of the pressure should be orthogonal to the shear expansion, $\pi^{ab}\sigma_{ab}=0$.  
\end{itemize} 
If both of these conditions are satisfied, with constant $\delta$ and $\lambda$, these must be zero.
We will therefore consider dynamical equations of state in the following subsection \ref{case1}.
However, it turns out that to find interesting cosmology one may have to violate one of the conditions above too.
We will consider this in the later subsection \ref{case2}. 
%nonlinear continuity equation, Eq.(\ref{cont_1}) into the identity $\rho_\Lambda=0$. 
%Let us then keep $\delta$ and $\gamma$ as free parameters describing the anisotropy of the $\Lambda$ term.
%Now there is a simple prescription which forces the isotropic pressure to compensate for the anisotropic part of the term:
%\be \label{visco}
%\hat{w} = -1-\gamma-\delta-\frac{1}{3}X.
%\ee
%The equations of state associated to the $\Lambda$-term can now be also different from $-1$. This was the model considered previously.

\subsubsection{Case I: Constant Isotropic Pressure.}
\label{case1}

One may satisfy both of the two requirements stated above with time-varying skewness parameters. We 
introduce a parameter $h$ and write
\bea
%w & = & -1 + h\theta(R-S), \label{r_w} \\ 
\delta  & = & \frac{h}{3}\theta(2S-R) \label{r_d}, \\ \label{r_w}
\gamma  & = & \frac{h}{3}\theta(S-2R)\label{r_g}.
\eea
Then the anisotropic stress becomes 
%as it appears in Eq.(\ref{decomp}) and is discussed in Section \ref{covariant}, becomes,  
\be
\pi^{a}_{\phantom{a}b} = diag\left[0,-S+R,S,-R\right]hH,
\ee
which indeed is, by construction, orthogonal to the metric shear 
\be
\sigma^{a}_{\phantom{a}b} = diag\left[0,R+S,-2R+S,R-2S\right] H/3. 
\ee
Since now also $w_X=-1$, both the energy density 
and the isotropic part of the pressure are constant. This prescription seems to be equivalent to the model introduced by Rodrigues \cite{Rodrigues:2007ny}, 
who also shows it is possible to associate such properties of the cosmological term with an infrared non-commutative property of the spacetime. 
The anisotropy arises from the deformation of the Poisson structure between the canonical momenta in such a way that the dynamics remain invariant 
under rescalings of the scale factors \cite{Rodrigues:2007ny}. The non-commutativity coefficients are then proportional to the parameter $h$ (which in 
our case has dimension of $1/\theta$). 

This would be difficult to apply for description of anisotropies in the present universe. The reason can be seen from skewness parameters
(\ref{r_d}) and (\ref{r_g}). Since they are proportional to the shear, the model does not support spontaneous generation of anisotropy. If $R$ 
and $S$ vanish in the early universe, they will stay zero even when the $\Lambda$ term becomes dominant. Observable deviation from $\Lambda$CDM could  
occur if the system turned out to be unstable: the $\Lambda$ term could in principle react to even perturbatively small anisotropies by amplifying them into 
direction-dependent background expansion. Let us briefly study this possibility. After algebraic manipulation one may write the evolution equations as 
follows
\bea
U' & = & -3U(U-1)(1+w_m), \\
R' & = & \frac{3}{2}\left(1-\frac{3\sigma^2}{\theta^2}\right)\left[-4h\theta SU-R(1+U-2h\theta U+(U-1)w_m)\right], \nonumber \\
S' & = & \frac{3}{2}\left(1-\frac{3\sigma^2}{\theta^2}\right)\left[ 4h\theta RU-S(1+U+2h\theta U+(U-1)w_m)\right], \nonumber
\eea
where the Hubble factor can be expressed as
\be
h\theta  =  \frac{\hat{h}}{\left[U(1-\frac{3\sigma^2}{\theta^2})\right]^{1/2}}
\ee
with the rescaled constant $\hat{h}$ which is now dimensionless, defined as $\hat{h} \equiv h\sqrt{24\pi G\rho_\Lambda}$.
The equation for $U'$ is of the same form as in $\Lambda$CDM cosmology. The solution is thus
\be \label{uell}
U(x) = \frac{1}{1+(\frac{1}{U_0}-1)e^{-3(1+w_m)x}},
\ee
where $U_0$ is the relative amount of vacuum density today. It is also obvious that $R=S=0$ is a solution with any value of $w$.
In fact the system has several fixed points. 
\begin{itemize}

\item The Kasner solutions
\be
R = \frac{S}{2} \pm \sqrt{9-\frac{3}{4}S^2}.
\ee
These correspond to the metric (\ref{kasner1}). Since we are not restricted here to a specific axisymmetric case, the Kasner exponents have only to satisfy
the relations (\ref{kasner2}). The solutions with different $-3<S<2$ correspond to these different exponents. Since for the Kasner solution matter is 
negligible, the value of $U$ is irrelevant (though formally should be set
either to $U=0$ or $U=1$). One quickly notices that this class of solutions
is unstable: the anisotropy decays and matter becomes eventually dominant.

%\item There exists an anisotropic $\Lambda$-dominated solution, with $U=1$, when 
%\be \label{ran1}
%-2.57433 \ge R \ge 0.720227 \| 1.40898 \ge R \ge 3.44512.
%\ee
%These numbers are approximate, since we have not found an analytic form for this fixed point. It is now also more convenient to express $\hat{h}$ in terms
%of $R$ than the other way around, though $R$ of course is the real dynamical variable while $\hat{h}$ is a parameter of the model:
%\be \label{ran2}
%\hat{h} = \frac{R}{R-2}\sqrt{\frac{9-(R-3)(R^2-6)R}{(R-1)^2}}.
%\ee  
%Thus, for each $\hat{h}$ there exist solution(s), if there are values of $R$ in the range (\ref{ran1}) that satisfy (\ref{ran2}).  
%There holds then a symmetry between $R$ and $S$:
%\be
%S = \frac{R}{R-1}, \quad R = \frac{S}{S-1}.
%\ee
%Numerical experimentation implies these solutions are not attractors.

\item The $\Lambda$ dominated isotropic solution
\be
R=S=0, \quad U=1.
\ee
If we then consider small perturbations $U=1+u$, $R=0 + r$, $S=0 + s$ about the $\Lambda$ 
dominated solution, we find that $$u \sim x^{-3(1+w_m)}, \qquad \textrm{while}
\qquad r,s \sim x^{-3 \pm i\sqrt{3}\hat{h}}.$$ Hence, the anisotropy decays, proportional to 
the inverse volume element. If anisotropy is inserted as an initial condition, the non-zero skewness parameters (\ref{r_d}) and (\ref{r_g}) only cause an 
oscillation about the usual diluting behaviour. 

\item The matter dominated isotropic solution
\be
R=S=0, \quad U=0.
\ee
One may also check that for the matter dominated solution, the perturbations go as
$$u \sim x^{3(1+w_m)}, \qquad r,s \sim x^{-3(1+w_m)/2}.$$ The negativity of the first exponent only signals that the fractional matter density is diminishing.  
The two last ones tell us that during matter domination the anisotropy decays too, though slower than during the $\Lambda$ domination, and to first order 
without  oscillations. 
\end{itemize}

For a cosmological evolution the relevant solutions are then the two last ones with $U=0$ and the other with $U=1$, and both having $R=S=0$. These will come out 
according to Eq.(\ref{uell}). The anisotropy is negligible by definition when the $\Lambda$ term can be neglected. Shear is not  
generated during the evolution of the universe, since anisotropic
perturbations about the usual solution decay. We have checked numerically that this 
behaviour is general and is not restricted to the asymptotic regimes of cosmological evolution.  

\subsubsection{Case II: Constant Anisotropic Pressure.}
\label{case2}

Let us now take a different approach, and keep $\delta$ and $\gamma$ as free parameters describing the anisotropy of the $\Lambda$ term.
The Eq.(\ref{continuity}) should still reduce to the identity $\dot{\rho}_\Lambda = 0$.
There is a simple prescription which achieves this by forcing the isotropic pressure to compensate for the anisotropic part:
\be \label{visco}
w_\Lambda = -1-\frac{1}{3}X.
\ee
The equations of state associated to the $\Lambda$-term can now be also different from $-1$. 
Also in such cases the vacuum energy density stays constant, but 
then the pressure becomes direction dependent, forcing eventually the universe to expand anisotropically 
when the cosmological term is significantly large.   
The evolution equations for the background variables then reduce to
\bea 
  U' & = & -3U(U-1)(1+w_m), \label{du} \\
  S' & = &  \label{ds}
       \frac{1}{6}
       \left(9 - R^2 + RS - S^2 \right)
       \left\{ S\left[-U\left(1+\frac{1}{3}X+w_m\right)+w_m-1\right]-6\gamma U \right\} \nonumber
\\
  R' & = &  \nonumber
       \frac{1}{6}
       \left(9 - R^2 + RS - S^2 \right)
       \left\{ R\left[-U\left(1+\frac{1}{3}X+w_m\right)+w_m-1\right]-6\delta U \right\}
\eea
The equation for $U'$ in (\ref{du}) is of the same form as in $\Lambda$CDM cosmology, and is given by Eq.(\ref{uell}). 
We have not found an analytic solution for the shears $R$ and $S$. 

For simplicity, let us consider the axisymmetric case that $S=\gamma=0$. We then find several fixed points. 
\begin{itemize}
\item The FLRW, 
\be
R=0, \quad U=0,
\ee
corresponding now to Eq.(\ref{flrw_s}).
It is easy to see that this point is always unstable. The matter domination will inevitably be followed 
by the cosmological constant dominated era, which is anisotropic whenever $\delta \neq 0$.
\item An anisotropic matter domination, 
\be \label{flrw_s2}
R=\pm 3, \quad U=0.
\ee
For dust matter this is always an unstable point.
\item An anisotropic $\Lambda$ domination I,
\be \label{alambda_s1}
R=\pm 3, \quad U=1.
\ee
For dust matter, this is a saddle point.
\item An anisotropic $\Lambda$ domination II,
\be \label{alambda_s2}
R=\frac{1}{2\delta}\left(3\pm\sqrt{9+36\delta^2}\right) , \quad U=1.
\ee
One may show that when $w_m=0$, this is always an attractor. More specifically, introducing small perturbations about this solution,
and constructing the matrix for the derivatives of these perturbations from the system (\ref{du}), the eigenvalues $q_u$, $q_r$ of the matrix then 
are 
\be
q_u = -3, \quad q_r = -\frac{3}{2\delta^2}\sqrt{1+4\delta^2}\left(\pm 1+\sqrt{1+4\delta^2}\right).
\ee   
Since these are negative regardless of the value of $\delta$, we know that the perturbations tend to decay.
\end{itemize}

A scenario then appears naturally, featuring a transition from the usual isotropic fixed point (\ref{flrw_s2}) to the anisotropic vacuum dominated 
point (\ref{alambda_s2}). This point coincides with the solution (\ref{alambda_s1}) when $\delta$ is very large. One notices that the asymptotic
value of the shear $R$ goes to $R \ra \pm 3$ when $\delta \ra \pm \infty$. We have numerically verified that the universe always evolves along this track
when begun from an isotropic initial stage. This then provides the possibility to generalize the standard $\Lambda$CDM cosmology in such a way that the universe 
features anisotropies at late times when the $\Lambda$-term begins to dominate. 

If the matter content is dust, the equation of state for the total fluid in this model is given by 
\be
w_{TOT} = -(1+\frac{X}{3})U,
\ee 
and the total sound speed squared is
\be \label{ccsound}
c_{sTOT}^2 = \frac{1}{9}X'\left(\frac{1}{\frac{1}{U}-1}\right) = \frac{\delta (R^2-9)U\left[18\delta U + R\left(3 + (3-\delta(3+2R))U\right)\right]}{81(U-1)},
\ee
assuming that $\delta$ is constant. One notes that this could be positive or negative. It vanishes when $\delta$ does: the usual $\Lambda$CDM universe
is ''silent'' in the sense that the CDM is pressureless, no perturbations propagate in the (isotropic) $\Lambda$ medium.
The anisotropy quantity $\an$ is now
\be
\an = \frac{2HR}{1 - \frac{1}{U} - \frac{2}{3}\delta R}.
\ee
It was expected that this is proportional to the shear in the expansion rate, $HR$.

\begin{figure}[ht]
\begin{center}
\includegraphics[width=0.7\textwidth]{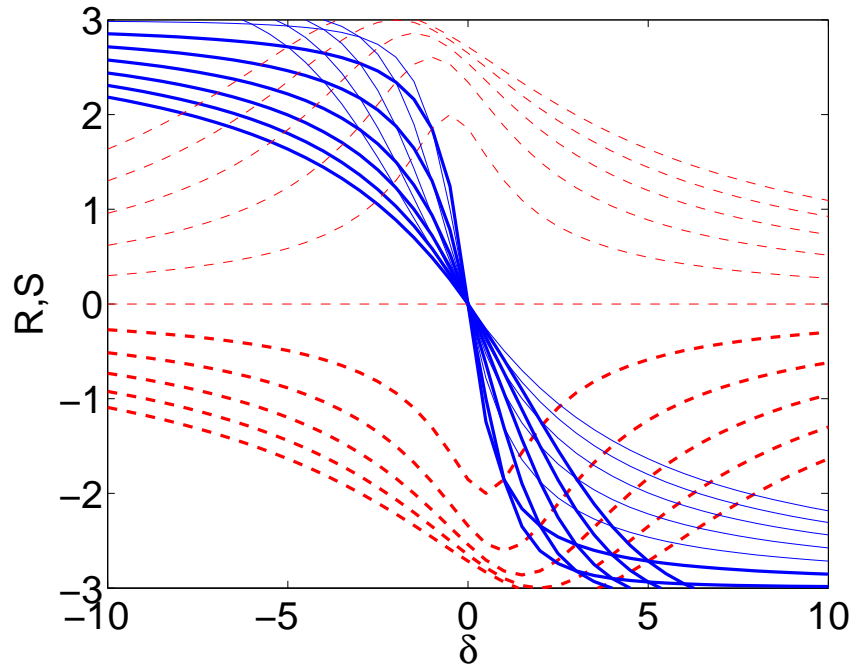}
\caption{\label{case2pic} The asymptotic state of the universe starting from an Einstein-deSitter 
stage for the anisotropic generalization of 
cosmological constant having constant $\delta$ and $\gamma$ (case II, section \ref{case2}). 
The $R$ (blue, solid lines) and $S$ (red, dashed lines) are given as function of $\delta$. 
The thick lines depict the cases $\gamma=1,2,3,4,5$, and the thin lines depict the case that $\gamma=0,-1,-2,-3,-4,-5$. The analytic solution 
(\ref{alambda_s2}) is recovered in the case $\gamma=0$. These curves show how this generalizes non-axisymmetric cases.}
\end{center}
\end{figure}

In the Figure \ref{case2pic} we show the  asymptotic state of the universe 
as a function of the skewness parameter $\delta$. In this numerically produced plot we also consider some cases where $\gamma$ does not equal zero, 
(though for simplicity we restricted to an axisymmetric case $S=\gamma=0$ in the analytic considerations). 
We return later to briefly consider the observational implications of this anisotropic generalization of $\Lambda$.

\section{Observational Bounds from an Anisotropic Background Expansion}
\label{bounds}

The part of this study focusing on homogeneous cosmology is ended by a comparison of the background dynamics of these models to the 
cosmological observations. In Section (\ref{basics})  we have made a phenomenological description of an anisotropic component
in terms of the skewness parameters $\delta$ and $\gamma$, Eq.(\ref{params}). 
%Such general parameterization is backed up by field theoretical considerations and explicit models of both inflation
%and dark energy were constructed. 
The question of its observational consequences is taken under investigation now.
We then focus on dark energy and an anisotropic background expansion. Observable effects which might have been imprinted during 
the early inflationary era, would appear to be more straightforwardly described as perturbations and thus not changing the FLRW 
predictions at zero order  \cite{Pontzen:2007ii,Ando:2007hc}. As we emphasise later, a proper treatment should take into account perturbations, in particular 
(the possibly anisotropic) initial conditions set at inflation \cite{Pitrou:2008gk}. However, as an attempt at a model-independent
first constraint on the model, it is useful to estimate the bounds ensuing from the background expansion. 
 
\subsection{CMB Anisotropies}
\label{cmb}

A possible anisotropy will leave its imprint in the CMB. To calculate the CMB spectrum in a background with a Bianchi I metric
Eq.(\ref{metric}), we begin from the geodesic equation for photons,
\be \label{geodesic}
\frac{d^2x^\mu}{d\lambda^2} + \Gamma^\mu_{\alpha\beta}\frac{d x^\alpha}{d\lambda}\frac{d x^\alpha}{d\lambda}
\ee
where the only nonvanishing connection coefficients are (summations are suppressed for Latin indices in this
section)
\be
\Gamma^0_{ii} = a_i \dot{a}_i, \quad \Gamma^{i}_{0i} = \Gamma^{i}_{i0} = \frac{\dot{a}_i}{a_i}.
\ee
We are interested in the redshift of photons coming from different directions.
The four vector $u^\mu = d x^\mu/d\lambda$ is subject to the null condition,
\be \label{null}
u^\mu u_\mu= 0 = -\left(\frac{d t}{d \lambda}\right)^2 + \sum_i a_i^2\left( P^i \right)^2,
\ee
where we call $u^i \equiv P^i$. Consider two photons, the first emitted at $t=t_e$ and the other a tiny
bit $\tau$ later at $t=t_e+\tau$. The difference of the two null conditions gives
\be \label{nulls}
u^0\frac{d\tau}{d\lambda} = \sum_i a_i\dot{a}_i \left(P^i\right)^2\tau(\lambda) + \mathcal{O}(\tau^2).
\ee
The definition of redshift is
\be \label{redshift}
1+z(\lambda_e) \equiv \frac{\tau(\lambda_r)}{\tau(\lambda_e)},
\ee
where $\tau(\lambda_r)$ is the time difference of the received signals.
Differentiating Eq.(\ref{redshift}) and  using Eq.(\ref{nulls}) yields
\be \label{diffz}
\frac{d \log{\left(1+z\right)}}{d\lambda} = - \frac{1}{u^0}\sum_i
a_i\dot{a}_i
\left(P^i\right)^2.
\ee
The $P^i$  can be obtained from the spatial components of the geodesic
equation (\ref{geodesic}),
\be \label{space_geo}
\frac{d P^i}{d\lambda} + 2\frac{\dot{a}_i}{a_i}P^i u^0 = 0.
\ee
We find $P^i \sim a_i^{-2}$. Since one can always rescale the scale
factors, let us fix $a_i(t_0)=1$ where $t_0$ indicates today.
We then write our initial $P^i$ as $P^i(t_0) = \hat{p}_i$.
Since one can reparameterize $\lambda$ without affecting the results, we
choose now such a normalization that $\sum_i \hat{p}_i^2 = 1$.
One may thus think about the unit vector $\hat{p}$ in terms of the angles
$(\hat{p}_x,\hat{p}_y,\hat{p}_z) = (\sin{\theta}\cos{\phi},\sin{\theta}\sin{\phi},\cos{\theta})$.
Thus we have managed to parametrize the photon
paths according to the angle they hit the observer today. Inserting this
solution for $P^i$ and $u^0$ from Eq.(\ref{null}) into the
evolution equation for the redshift (\ref{diffz}), one finds that it can
be solved to yield
\be \label{z_p}
1+z(\hat{{\bf p}}) = \left( \sum_i
\frac{\hat{p}^2_i}{a_i^2}\right)^\frac{1}{2}.
\ee
Since the sum is equal to
\be
\sum_i \frac{\hat{p}^2_i}{a_i^2} = \frac{1}{a^2}\left[\sin^2\theta\left(\cos^2\phi + \frac{a^2}{b^2}\sin^2\phi\right) +
\frac{a^2}{c^2}\cos^2\theta\right],
\ee
it is easy to see that one can rewrite the result as
\be \label{z_ellips}
1+z(\hat{{\bf p}}) =
\frac{1}{a}\sqrt{1 + \hat{p}_y^2 e_y^2 + \hat{p}_z^2e_z^2}
\ee
in terms of the eccentricities
\be \label{ellips}
e_y^2 = \left(\frac{a}{b}\right)^2-1, \quad
e_z^2 = \left(\frac{a}{c}\right)^2-1.
\ee
The temperature field is determined by this redshift by the relation
\be
T(\hat{{\bf p}}) = \frac{T_*}{1+z(\hat{{\bf p}})},
\ee
and the spatial average is $$4\pi\bar{T} = \int d\Omega_{\hat{{\bf p}}}T(\hat{{\bf p}}).$$
The last scattering temperature $T_*$ does not depend on the direction, but the photons coming from different directions will be redshifted different amounts.  
The anisotropy field is then
\be
\frac{\delta T(\hat{{\bf p}})}{\bar{T}} = 1- \frac{T(\hat{{\bf p}})}{\bar{T}}.
\ee
The coefficients in the spherical expansion of this anisotropy field are called
$a_{\ell m}$, and due to orthogonality of spherical harmonics $Y_{\ell m}$, are given by
\be
a_{\ell m} = \int d\Omega_{\bf p} \frac{\delta T(\hat{{\bf p}})}{\bar{T}} Y^*_{\ell m}.
\ee
The multipole spectrum can be described by
\be \label{multipole}
Q_\ell = \sqrt{\frac{1}{2\pi}\frac{\ell(\ell+1)}{(2\ell+1)}\sum_{m=-\ell}^{\ell}|a_{\ell m}|^2}.
\ee
One may expand the redshifts (\ref{z_ellips}) in the eccentricities (\ref{ellips}).
Then one notes that the $a_{\ell m}$ will be real (since there is only even dependence on
the polar angle in the anisotropy field and the imaginary parts of the $e^{i m\phi}$ integrate to zero), and that
for all odd $\ell$ the $a_{\ell m}$ will vanish (since only even powers of the azimuthal appear in the expansion).
To  first order in $e_{x,y}^2$, $$\bar{T} = aT_*[1-\frac{1}{6}(e_z^2+e_y^2)],$$ and in addition to the monopole,
there is only the quadrupole
\bea 
a_{20}=\frac{1}{3}\sqrt{\frac{\pi}{5}}\left(2e_z^2-e_y^2\right), \qquad
a_{21}=a_{2-1}= 0, \qquad a_{22}=a_{2-2}= -\sqrt{\frac{\pi}{30}}e_y^2 \nonumber
\eea
which implies
\bea
\label{quad}
Q_2 = \frac{2}{5\sqrt{3}}\sqrt{e_z^4+e_y^4-e_z^2e_y^2}.
\eea
The observed value of this is $Q_2(obs) \approx  5.8 \, \, 10^{-6} $, while the standard
concordance model predicts $Q_2(iso) \approx  1.2985  \, \, 10^{-5}$. It has been suggested in
previous works also that this discrepancy could be explained by an ellipsoidality of
the universe  \cite{Campanelli:2006vb,BeltranJimenez:2007ai}. This would require that the
anisotropy of the
background is suitably oriented with respect to the intrinsic quadrupole and cancels its power to
sufficient amount. For any orientation then, we should have $Q_2 \lesssim  2.7209  \,  \, 10^{-5}$ to
be consistent with observations taking into account the cosmic variance.
The constrains this implies on the dark energy equation of state parameters are rather tight.

One should, however, keep in mind that we have only studied the Bianchi I type background, which leaves a very simple pattern
on the CMB, and assumed that to be somehow superposed with the anisotropies of the standard perturbed FLRW universe. This is of 
course a gross description. In a more realistic treatment, one should study the perturbations in matter in the anisotropic 
background. This is not a straightforward problem and will be tackled in
section \ref{covariant1}. We now focus only on the background anisotropy and 
assume that at least the order of magnitude of the limits one derives from it reflect correctly the properties of the universe.  

In more general models, e.g. with time-varying $\delta$ and $\gamma$, one could allow more anisotropy. It is in 
principle possible for arbitrarily anisotropic expansion to escape detection from CMB as long as the expansion rates evolve in such a way that $e_z=e_y=0$. In 
other words, the quadrupole vanishes, if each scale factor has expanded - no matter how anisotropically - the same amount since the last scattering 
\footnote{We emphasize that this is only true for the CMB pattern due to the background anisotropy. In the next Section of the article we consider 
inhomogeneous perturbations, which in general have anisotropic signatures also in the case that they vanish at the background level.}.

Generating these effects at low redshift has an advantage that
it relaxes constraints which would otherwise come from the CMB polarization \cite{Pontzen:2007ii} and could be strong for a given
temperature anisotropy in isotropizing models because of the significant polarization anisotropy at last scattering.
However, anisotropic dark energy could evade this since the optical depth to $z \sim 1$ is very small. More specifically, polarization of the 
CMB photons is generated by the Thomson scattering process on the free electrons. When the electron distribution has quadropole
anisotropy, the scattered radiation will have net polarization. Thus it is clear that an anisotropy of the expansion will in principle
always have imprints on the polarization pattern. However, most of the photons we see in the CMB were last scattered at $z \sim 1100$.
Any model which predicts large anisotropies at such redshifts could thus easily be ruled out. At smaller redshifts, the effects of anisotropy
on polarization are only secondary, since the photons do not scatter, but their energy distribution may be slightly modified. At the reionization epoch
one expects again some photons to scatter, but the impact this has on CMB is restricted to the largest scales and suppressed 
by several orders magnitude. Thus it seems easier to explain the observed anomalies in the temperature spectrum without running into problems
with unobserved anomalies in the polarization spectrum by anisotropies occurring in the later rather than in the earlier universe.

\subsection{SNIa Luminosities}
\label{supernova}

Anisotropy in the acceleration rates would be in principle seen as rotationally non-invariant
luminosity distance - redshift relationships for the SNIa. The main CMB constraints on the present model ensue from
the quadrupole moment subject to effects of cosmic variance, and at least in specific models, the constraints can be loose or 
nonexistent. It is then interesting to see if one could
limit the anisotropy with the present data on the nearby supernovae. More accurate SNIa data
will become available in the near future, notably from the SNAP experiment\footnote{see http://snap.lbl.gov/.}, but we find that 
useful constraints can already be derived. The luminosity-redshift relationship of the SNIa can be used to probe the possible
anisotropies in the expansion history. This is a complementary probe to the CMB quadrupole, since the SNIa objects
are observed at the $z < 2$ region, whereas CMB comes from much further away at $z \sim 1000$. 
The luminosity distance as a probe of anisotropies has been considered in \cite{Bonvin:2005ps,Schwarz:2007wf}.

The spatial geodesic eq.(\ref{space_geo}) tells us that the direction ${\bf \hat{p}}$ of a photon that is coming towards us
is constant. This reflects the orthogonality property of the Bianchi I model, in contrast to tilted models where
the geodesics would involve rotation effects. For the null geodesics we have then
\be
-dt^2 = \left(\hat{p}_x^2a^2+\hat{p}_y^2b^2+\hat{p}_z^2c^2\right)dr^2,
\ee
where $r^2 = x^2+y^2+z^2$. The conformal distance is computed as $\int dr$.
The luminosity distance at the redshift $z$ in the direction $\hat{p}$ is thus given by
\be \label{lumi}
d_L(z,{\bf \hat{p}}) = (1+z)
\int_{t_0}^{t(z)}\frac{dt}{\sqrt{\hat{p}_x^2a^2+\hat{p}_y^2b^2+\hat{p}_z^2c^2}}.
\ee
To test this prediction with the data, we apply the formula (\ref{z_ellips}) for each
observed redshift of a supernova and match its distance modulus $\mathcal{M}$ inferred from the observation
to the one computed from Eq. (\ref{lumi}) by
\be
\mathcal{M} = 5\log_{10}\left(\frac{d_L(z,{\bf \hat{p}})}{10 \rm{pc}}\right).
\ee
We have to then take into account also the angular
coordinates of each individual supernovae in the sky which fix ${\bf \hat{p}}$ for each object.
In our analysis we use the GOLD data set \cite{Riess:2006fw}, which consists of five
subsets of data\footnote{We find the angular coordinates of each of the 182 GOLD supernovae
partly from  \cite{Riess:2006fw} and  \cite{Astier:2005qq} and from
http://cfa-www.harvard.edu/ps/lists/Supernovae.html.}. We marginalize over the directions
in the sky by integrating over the likelihood-weighted $\chi^2$ over three Euler angles and normalizing.
Similarly we also marginalize over the present value of the Hubble constant. Taking these cosmologically
irrelevant parameters into account makes the computations much heavier but is necessary.  
\begin{figure}[ht]
\begin{center}
\includegraphics[width=0.49\textwidth]{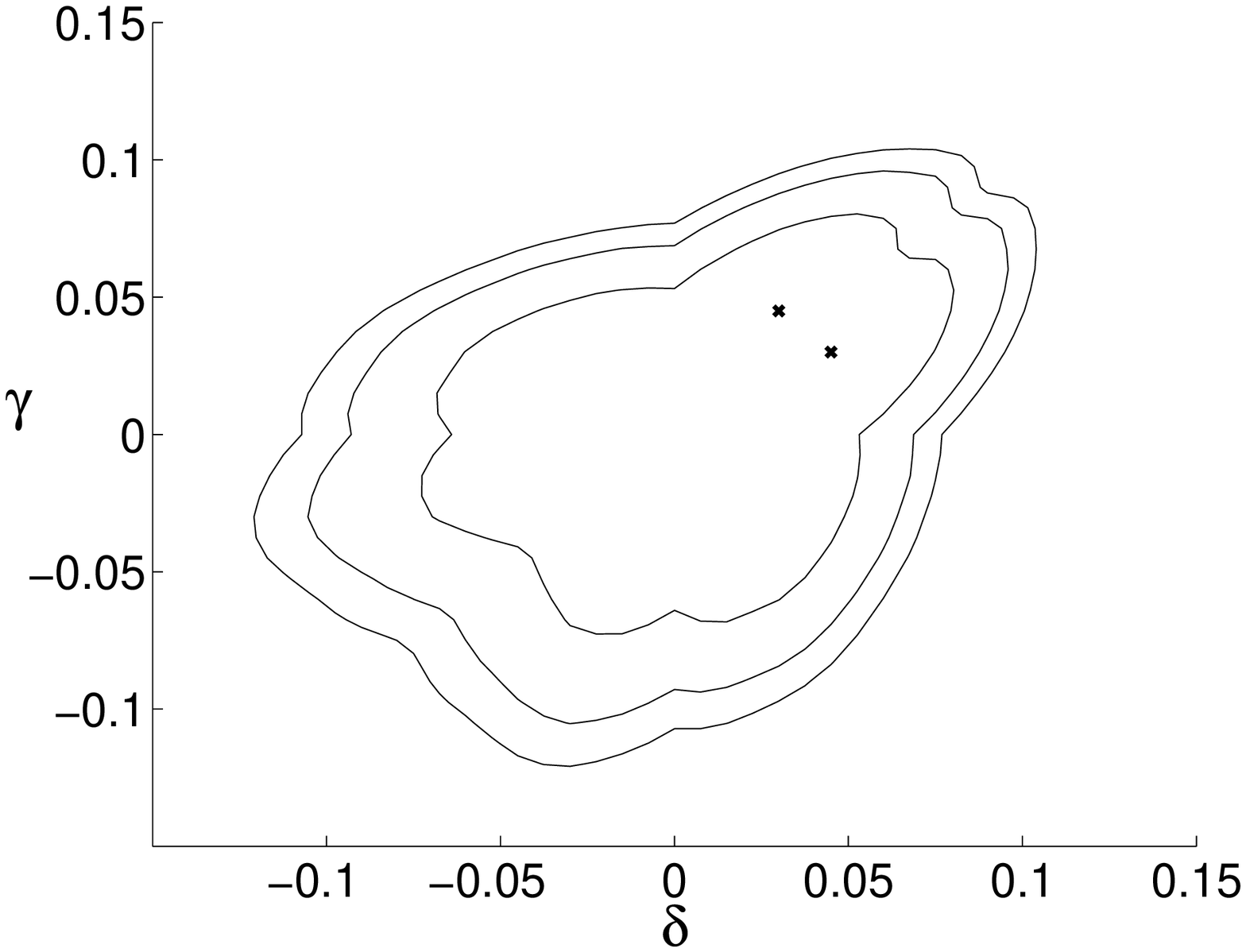}
\includegraphics[width=0.49\textwidth]{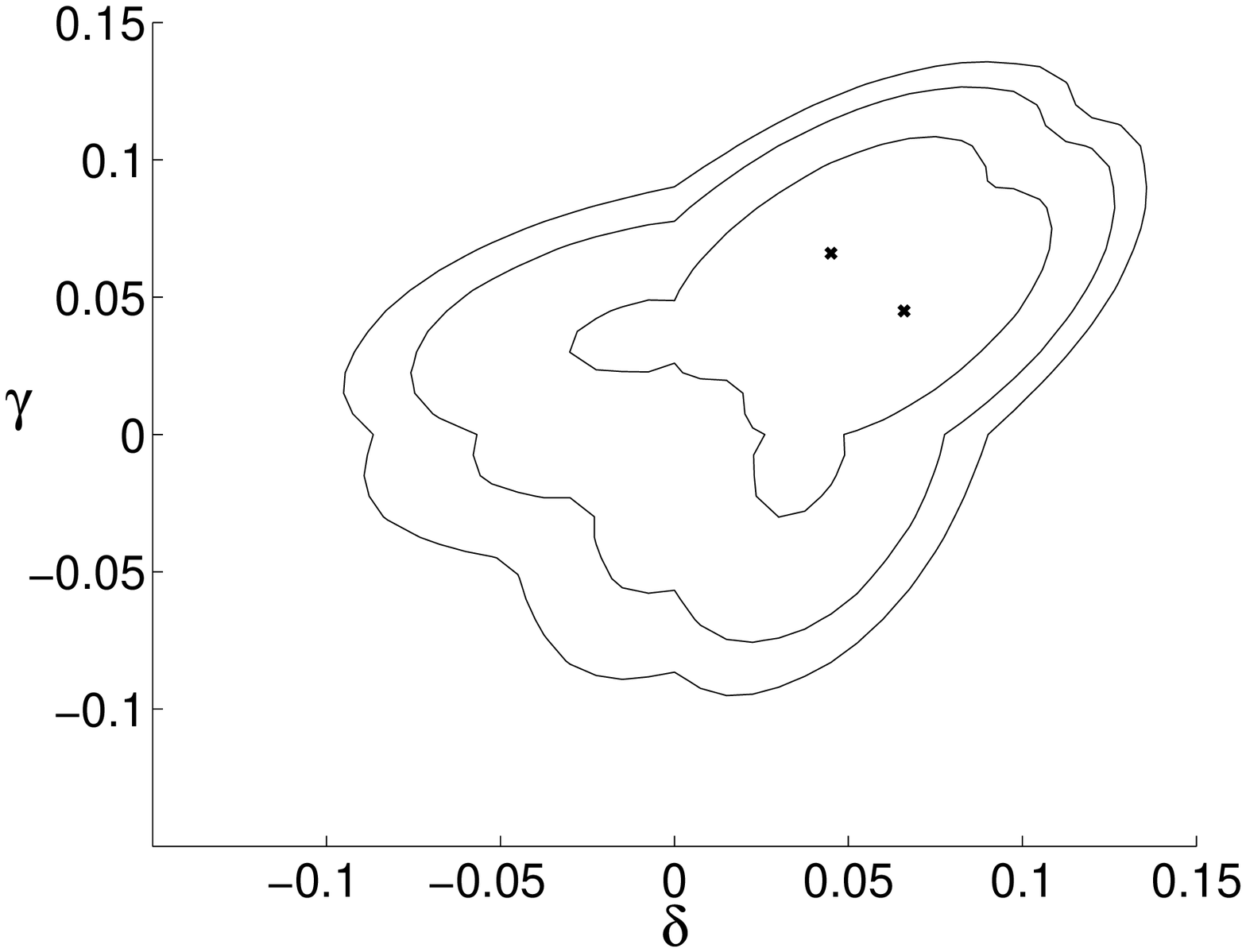}
\caption{\label{eoslims} Limits on the skewness of dark energy arising from the SNIa data
when $\Omega_m=0.3$. In the LHS, $w=-1$ and RHS $w=-1.2$. The contours correspond to
68.3, 90, and 95.4 percent confidence limits. The crosses mark the location of the best-fit models (there
are two in each figure since, because we marginalize over the direction of the coordinates, there is a symmetry
of projections about the $\delta+\gamma=0$ line).}
\end{center}
\end{figure}
The results are summarized in Fig. \ref{eoslims}. The best-fit anisotropic models
are only slightly preferred over the $\Lambda$CDM, the difference being $\Delta \chi^2 \approx 1$.
On the other hand, the SNIa data allows skewness in dark energy to an interesting degree.

\begin{figure}[ht]
\begin{center}
\includegraphics[width=0.49\textwidth,height=0.55\textwidth]{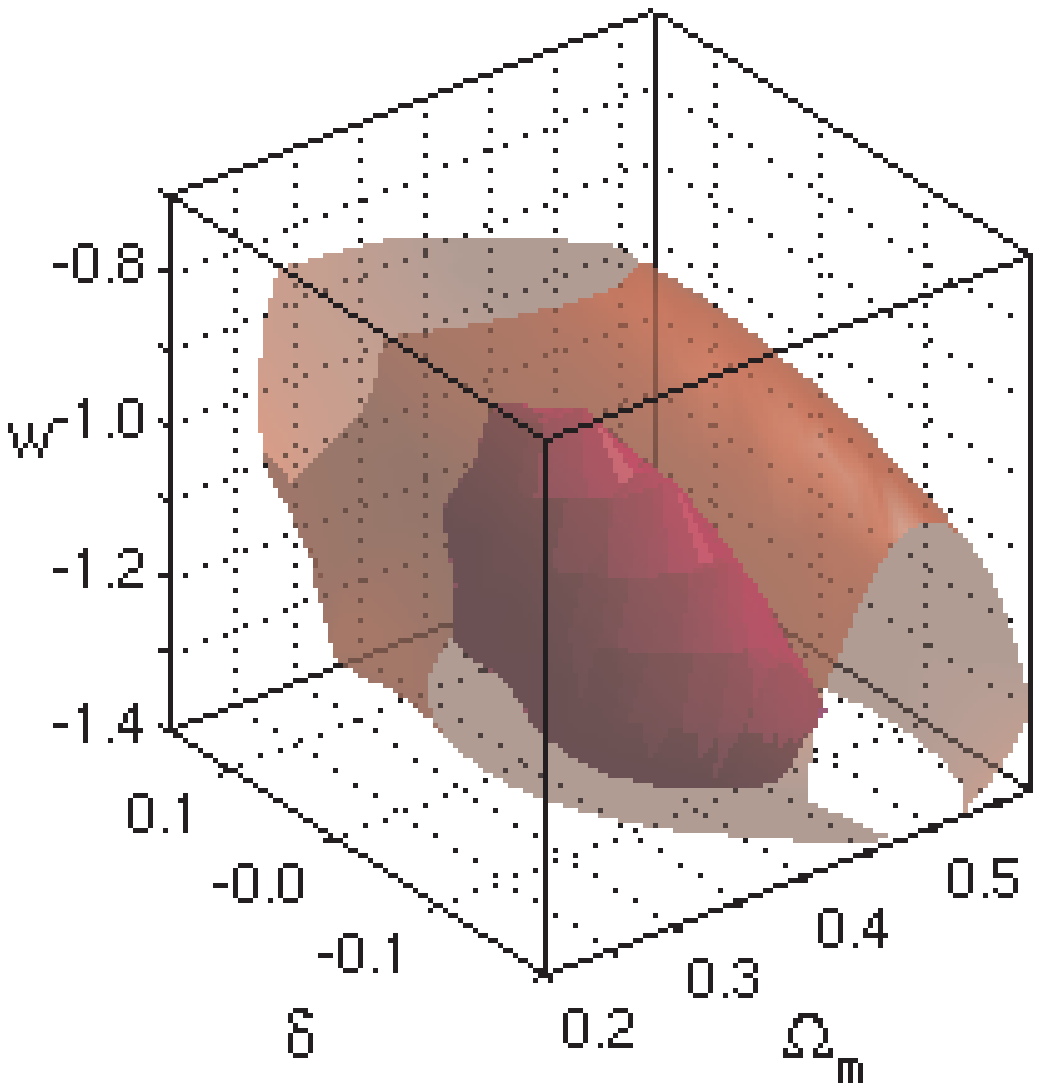}
\includegraphics[width=0.49\textwidth,height=0.55\textwidth]{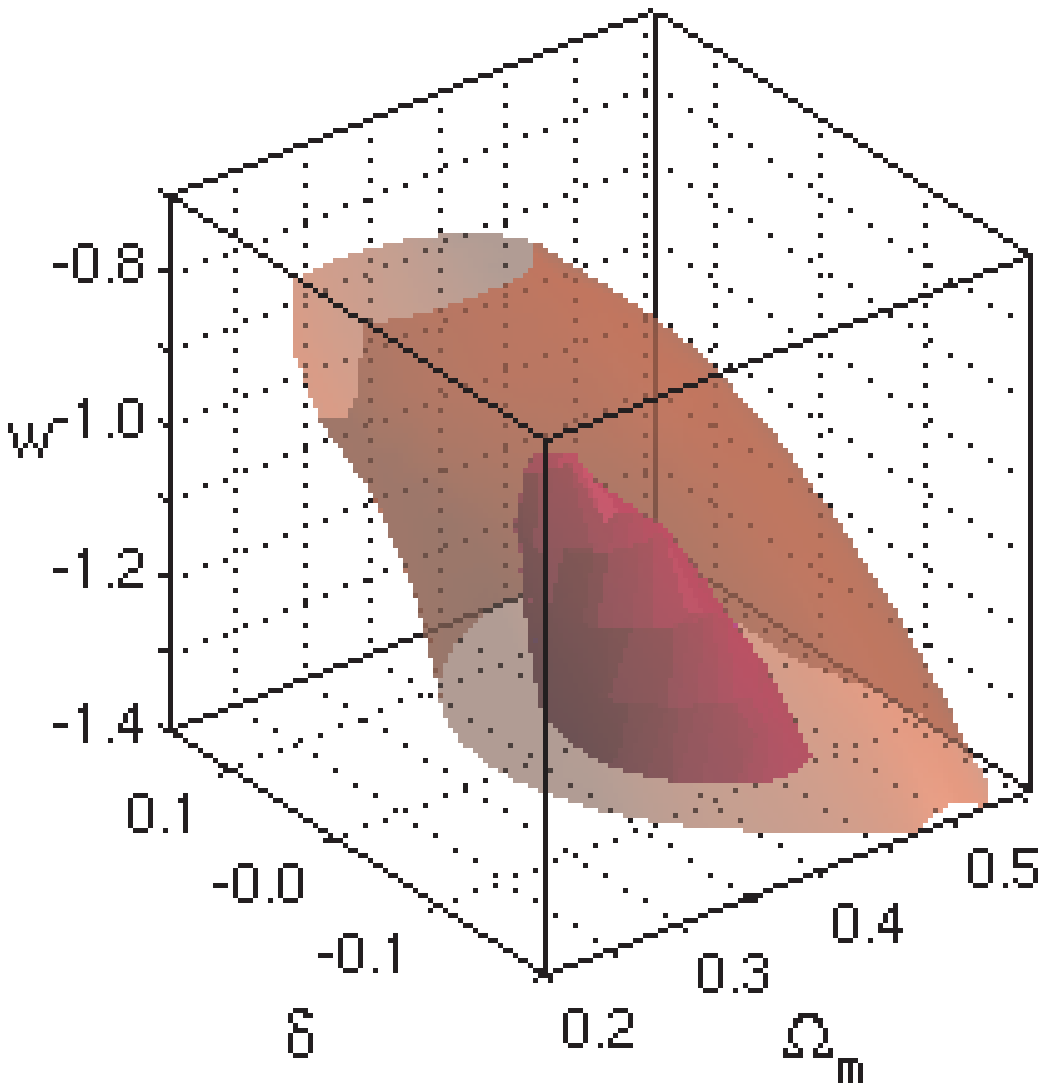}
\caption{\label{axislims1} Constraints arising from the SNIa data in the axisymmetric case. Inside the darker
isosurfaces, the fit is as good as in the $\Lambda$CDM model, $\chi^2 < 158$. Inside the lighter shaded
isosurfaces, one has $\Delta\chi^2 < 8.02$. In the left hand side figure, $\delta = \gamma$ and thus $w$ is
the equation of state along the symmetry axis. In the right hand side figure, $\gamma=0$ and thus $w + \delta$ is
the equation of state along the symmetry axis. One notes that larger skewness $\delta$ would typically be
compatible with the SNIa data for ''supernegative'' equations of state $w<-1$ and large matter densities.}
\end{center}
\end{figure}

In Fig. \ref{axislims1} we illustrate the SNIa constraints in the axisymmetric cases.
Larger skewness $|\delta|$ would typically be compatible with the SNIa data
for equations of state $w<-1$ and large matter densities.
This means that  even if the CMB formed isotropically at early time,
it could be distorted by the acceleration of the later universe in such a way
that it appears to us anomalous at the largest scales. The future SNIa
data, with considerably improved error bars
%\footnote{These are expected in particular from the SNAP experiment, see http://snap.lbl.gov/.} on $d_L(z,{\bf \hat{p}})$ 
can be used to rule out this possibility, and to distinguish whether the possible statistical anisotropy was
already there at last scattering or whether it is due to dark energy.

\section{Inhomogeneous Cosmology}
\label{covariant1}

\subsection{Covariant Perturbations}
\label{covariant}

In the previous sections we have investigated the behaviour of an anisotropic universe which is homogeneous.  
%Firstly, there seems to several significant hints of a cosmic anisotropy: The CMB temperature field seems asymmetric \cite{Eriksen:2003db}; there also 
%appears to be some unexpected correlations between the lowest multipoles of the spectrum \cite{Land:2005ad}; and recent analysis have shown that these effects in 
%fact extend to higher multipoles \cite{Samal:2007nw}. Secondly, the universe seems to be accelerating, and this requires some exotic 
%energy source. There has been several proposals for a possible connections between dark energy and the anomalous CMB.  
%Vector fields, spatial fields in form of a triad \cite{Armendariz-Picon:2004pm}, time-like massive field \cite{Boehmer:2007qa}, 
%or general fields with a possible Gauss-Bonnet coupling \cite{Koivisto:2007bp} have been proposed. Parameterization of a 
%generalized fluid including a shear anisotropy of a physically motivated viscous form has been investigated in detail taking into 
%account the effects at the perturbative in a FLRW universe \cite{Koivisto:2005mm,Mota:2007sz}.
Clearly this cannot be a realistic description of the universe, though it suffices as a good approximation when considering
the expansion dynamics relevant to e.g. the supernovae luminosity distances. When considering e.g. the formation of galaxies and
other structures in the universe, one has to allow perturbations about the homogeneous cosmological solution. 
Perko {\it et al} \cite{Perko:1972cs} pioneered the study of galaxy formation in anisotropic cosmologies.
They discovered that in an anisotropic background, the gravitational waves can couple with the density modes.
One of their results was that the shear can enhance the growth rate of matter perturbations.
Later Bianchi I perturbations have been also considered within the coordinate approach to cosmological perturbations. 
The perturbation equations have been written quite explicitly in a gauge-ready form \cite{Noh:1987vk}, including
then more general sources. However, most studies have considered dust cosmologies \cite{Miedema:1992wb,Dimastrogiovanni:2008ua}.
In view on inflationary models, limiting matter sources to a slowly rolling scalar field is of course also of a particular interest.
Indeed, very recently there have been studies of perturbations in anisotropic inflation 
which allow an anisotropically expanding background \cite{Gumrukcuoglu:2007bx,Pereira:2007yy}. 
%There are several recent developments of the evolution of anisotropic universes at the perturbative level \cite{Pontzen:2007ii,Ando:2007hc}.  
Considering late accelerating cosmologies, perturbation theory for elastic dark energy has been developed \cite{Battye:2007aa}. Yet this has only been done in the 
statistically isotropic framework, though the models would allow an anisotropic generalization \cite{Battye:2006mb}.
%In this section we go beyond those studies and investigate in detail the evolution and behaviour of inhomogeneities produced within these Anisotropic
%models. 
%Analysis of the nonlinear evolution of anisotropic power shows that the anisotropies are only slightly suppressed at scales corresponding to quasi-linear 
%evolution \cite{Ando:2007hc}.
In this paper we work within the covariant formalism. Therefore our notation follows most closely Dunsby's study of
Bianchi I perturbations \cite{Dunsby:1993fg} in the covariant approach, which we slightly generalize from the point of view
of possible non-isotropizing cosmologies, thus taking into account imperfect sources \cite{Dunsby:1991vv}.
Maartens and Tsagas have considered imperfect sources in the form of magnetic fields in the covariant approach in detail \cite{Tsagas:1999ft,Tsagas:1999tu}.
We will omit most of the details of the covariant formalism and refer to, for instance the
recent review of large-scale structure in relativistic cosmology discussing also more general spacetimes
than the FLRW model \cite{Tsagas:2007yx}.

\subsubsection{Definitions.}

In a homogeneous background universe it is then convenient to consider variables which are defined as spatial gradients of 
some physically interesting quantities. These should then, by definition, count as perturbations. Since these gradients vanish in the background, the variables 
are by construction gauge-invariant. The basic variables we employ are defined as
\be
D_a = \frac{\ell}{\rho}\nabu_a\rho, \quad Y_a = \nabu_a p, \quad Z_a = \ell\nabu_a \theta, \quad \Omega_a = \ell\nabu^b\omega_{ab}
\ee
The first two of these are fluid quantities related to the energy density and pressure. $Z_a$ is the gradient of the 
expansion and $\Omega_a$ is the rotational perturbation.

In addition to these, we define some quantities related to the imperfect properties of the fluid:
\bea
\Pi_a & = & h^c_{\phantom{c}a}\pi^b_{\phantom{b}c;b}, \\ 
\Xi_a & = & h^b_{\phantom{b}a}\left(\pi_{cd}\sigma^{cd}\right)_{;b}, \\
\Sigma_a & = & \ell \nabu_a \sigma.
\eea
It is also useful to define two scalars, the divergences of the acceleration and of the heat transfer,
and two vectors, the spatial gradients of these as
\be \label{defs}
A= a^a_{\ph{a};a}, \quad A_a = \nabu_a A, \qquad
Q = q^a_{\ph{a};a}, \quad Q_a = \nabu_a Q.
\ee
From these definitions it follows that
\be \label{aux}
\nabu^a q_a = Q - a_a q^a, \quad \nabu^b\pi_{ab} = \Pi_a - h_a^{\ph{a}b} a^c \pi_{bc}.
\ee
Note that our conventions are slightly different from the Tsagas {\it et al}  \cite{Tsagas:2007yx}, and
when comparing the results the relations (\ref{aux}) are useful.

For a fluid with an equation of state $p=p(\rho,s)$, where $s$ is the entropy, we decompose the spatial gradient
of the pressure as
\be
\ell Y_a = c_s^2\rho D_a + pE_a, \quad E_a = \frac{\ell}{p}\left(\frac{\partial p}{\partial s}\right)\nabu_a s,
\ee
where the adiabatic sound speed is, at the zeroth order, related to the equation of state $w \equiv p/\rho$ by
\be
\dot{w} = (w-c_s^2)(1+w)(\theta + \an).
\ee
We have defined an auxiliary variable $\an$ to characterize the anisotropy,
\be
\an \equiv \frac{\pi_{ab}\sigma^{ab}}{\rho(1+w)}.
\ee
Using this variable will considerably simplify the following equations.

Note that the dot corresponds to the derivative along the flow as stated in Eq.(\ref{timed}). It corresponds to the derivative with respect to
cosmic time $t$ only for scalars.

\subsubsection{Nonlinear Equations.}
\label{nonlinear}

The Raychaudhuri equation is
\be \label{ray}
\dot{\theta} + \frac{1}{3}\theta^2 + 2\left(\sigma^2-\omega^2\right) - A + \frac{1}{2}\left(\rho+3p\right) = 0.
\ee

By projecting the covariant divergence of the stress-energy along and orthogonal to
$u^a$, one obtains, respectively, the two conservation equations
\be \label{cont1}
\dot{\rho} + \left(\rho+p\right)\theta + (1+w)\rho \an + Q = 0,
\ee
\be \label{cont2}
\left(\rho + p\right) a_a + Y_a + \Pi_a + \mathcal{Q}_a = 0,
\ee
where we have defined
\be
\mathcal{Q}_a = \dot{q}_a + \sigma_a^{\phantom{a}b}q_b + \frac{4}{3}\theta q_a.
\ee
The shear evolution equation can be written as
\be \label{shear1}
\dot{\sigma}_{<ab>}-\nabu_{<a}a_{b>}+\frac{2}{3}\theta\sigma_{ab}+\sigma_{c<a}\sigma_{b>}^{\phantom{a}c}+E_{ab}+\frac{1}{2}\pi_{ab} = 
-a_{<a}a_{b>}
+\omega_{<a}\omega_{b>}.
\ee
We will be able to drop away the RHS of this equation at the linearized level. The $<>$ means now the traceless symmetric par. $E_{ab}$ is the electric 
part of the Weyl tensor. This far our equations have been exact and thus very general but not very solvable in practice. 

\subsubsection{Linearized Evolution of Density.}
\label{linear}

We now linearize the equations, supposing an anisotropic but homogeneous background. We assume that the heat flux and vorticity 
vanish at the zero order. All the possible spatial gradients vanish in the background, since it is homogeneous. However, we 
allow anisotropy both in the metric, $\sigma_{ab}$, and in the fluid $\pi_{ab}$. Note that in this subsection we do not yet 
specify the  type anisotropy, and thus the following equations apply in any Bianchi type spacetime. 

We find that the overdensity evolves as
$$
\dot{D}_a = w\theta D_a -(1+w)Z_a-\sigma^c_{\ph{c}a}D_c
+ (1+w)\an\left(D_a - \ell a_a\right) + \frac{\ell}{\rho}\left(\theta\Pi_a-\Xi_a\right)
+ \frac{\ell}{\rho}\left(\theta\mathcal{Q}_a - Q_a\right)
%\left(\dot{q}_a+\frac{4}{3}\theta q_a + \sigma_a^{\ph{a}b}q_b\right)
$$
where $w \equiv p/\rho$. The evolution equation for the gradient of the expansion scalar is
\be
\dot{Z}_a = -\frac{1}{2}\rho D_a-\frac{2}{3}\theta Z_a - \sigma^c_{\ph{c}a}Z_c - 3 \ell\sigma^2 a_a
+ \ell A_a - 4\sigma\Sigma_a + \frac{3}{2}\ell\Pi_a + \frac{3}{2}\ell\mathcal{Q}_a
%\left(\dot{q}_a + \sigma_a^{\ph{a}b}q_b + \frac{4}{3}\theta q_a\right).
\ee
From the continuity equation (\ref{cont2}) we see that $A_a$, as defined in (\ref{defs}), may be given as
$$
A_a  =  -\frac{1}{\rho(1+w)}\left(\nabu^2 Y_a + \nabu_a\nabu^b\nabu^c\pi_{bc}
+ \nabu_a\nabu^b\mathcal{Q}_b\right) - 2c_s^2\theta \Omega_a 
$$
where one can show that
\be
\nabu_a\nabu^b\mathcal{Q}_b = (\nabu_a Q)^\bullet + 2\theta\nabu_a Q + \sigma^{b}_{\ph{b}a}\nabu_b Q
+ 2\sigma^{bc}\nabu_a\nabu_b q_c.
\ee

The evolution equation for the density gradients follows then by differentiating the first order equation 
and using all the previous formulas. We find 
\bea \label{d_evol} 
\ddot{D}_a & + & \left[(\frac{2}{3}+c_s^2-2w)\theta
- (1+2w)\an \right]\dot{D}_a
+ 2\sigma_a^{\ph{a}b}\dot{D}_b
\\ \nonumber
& - &
\left[(\frac{1}{2}+4w-\frac{3}{2}w^2-3c_s^2)\rho
+ (2w-6c_s^2)\sigma^2 + c_s^2 \nabu^2 \right]D_a
\\
&+& \left[(\frac{2}{3}-2w+c_s^2)\theta
- (1+2w)\an\right]\sigma_a^{\ph{a}b}D_b
+ \sigma_a^{\ph{a}b}\sigma_b^{\ph{b}c}D_c 
- \pi_a^{\ph{a}b}D_b
\nonumber
\\\nonumber
&-&
\Big[(1+c_s^2+w)\dot{\an} +
\left(\dot{c}_s^2 + \frac{1}{3}\theta (2 + c_s^2 (-1 + 3c_s^2 - 3w) + 5 w)\right)\an
\\
&+& c_s^2(c_s^2-w)\an^2\Big]D_a
-  \left[w(\nabu^2-3\sigma^2 + \dot{\an})
+ (\frac{5}{3}w-c_s^2)\theta\an
+ (w-c_s^2)\an^2\right]E_a\nonumber
\\
&-& w\an(\dot{E}_a + \sigma_a^{\ph{a}b}E_b)
+ 2(1+w)c_s^2\theta\Omega_a
-   4\sigma(1+w)\Sigma_a = S^{(q)}_a + S^{(\Pi)}_a.\nonumber
\eea
The right hand side of Eq.(\ref{d_evol}) vanishes if we consider a case that the imperfect perturbations can be neglected. 
The sources are due to the heat flux perturbations as
\bea \label{s_hf} \frac{\rho S^{(q)}_a}{\ell}  &=&  -   \dot{Q}_a -  \left[(2+c_s^2)\theta + (1 + c_s^2) \an^2\right]Q_a-  \sigma_{a}^{\ph{a}b}Q_b
\\ 
  & + &  \nabu_a\nabu^b\mathcal{Q}_b+  \left(\theta + \an \right)\dot{\mathcal{Q}}_a
+   \an\sigma_a^{\ph{a}b}\mathcal{Q}_b
 \nonumber \\ &+&  \left[(3w-2)\sigma^2 + \rho+ \dot{\an}+ (3 + c_s^2 + w)\theta\an+ (1+c_s^2)\an^2\right]\mathcal{Q}_a \nonumber
\eea
%\begin{multline} \frac{\rho S^{(q)}_a}{\ell}  =  -   \dot{Q}_a -  \left[(2+c_s^2)\theta + (1 + c_s^2) \an^2\right]Q_a-
%\sigma_{a}^{\ph{a}b}Q_b +  \nabu_a\nabu^b\mathcal{Q}_b + \left(\theta + \an \right)\dot{\mathcal{Q}}_a
%\\ +  \left[(3w-2)\sigma^2 + \rho+ \dot{\an}+ (3 + c_s^2 + w)\theta\an+ (1+c_s^2)\an^2\right]\mathcal{Q}_a
%+  \an\sigma_a^{\ph{a}b}\mathcal{Q}_b
%\end{multline}
The sources from perturbations of the anisotropic stress are the following
\bea \label{s_pi} \frac{\rho S^{(\Pi)}_a}{\ell}    &=&   \nabu_a\nabu^b\nabu^c\pi_{bc} - \dot{\Xi}_a - \left[(\frac{4}{3} + w)\theta +  (1 + w)\an\right]\Xi_a
\\ 
 & + & (\theta+\an)\left(\dot{\Pi}_a + \sigma_{a}^{\ph{a}b}\Pi_b\right)   
\nonumber \\ &+&
\left[3(1-w+c_s^2)\rho + 3c_s^2\sigma^2 + \dot{\an} + (3+2c_s^2)\theta\an + (1+c_s^2)\an^2\right]\Pi_a. 
\nonumber
\eea
%\begin{multline} \frac{\rho S^{(\Pi)}_a}{\ell}   =  \nabu_a\nabu^b\nabu^c\pi_{bc} - \dot{\Xi}_a - \left[(\frac{4}{3} + w)\theta
%+  (1 + w)\an\right]\Xi_a
%+ (\theta+\an)\left(\dot{\Pi}_a + \sigma_{a}^{\ph{a}b}\Pi_b\right) \\  +
%\left[(3+c_s^2)\sigma^2 + \frac{3}{2}(3-w+4c_s^2)\rho + \dot{\an} + (3+2c_s^2)\theta\an + (1+c_s^2)\an^2\right]\Pi_a
%\end{multline}
We cannot derive the general evolution equations for these terms. Their behaviour depends completely on the nature of the fluid.
In fact this is true also for the other quantities, but then we have a simpler and conventional way of parameterizing the 
unspecified properties in terms of $w$, $c_s^2$.    

\subsubsection{Linearized Evolution of Shear.}

In addition to the unspecified matter part, the shear $\sigma$ and its gradient $\Sigma_a$ appear in our evolution equations. So we should find a way to 
solve for them too. From the shear evolution equations (\ref{shear1}) we easily find that now
\be \label{sigmadot}
\dot{\sigma} = \frac{1}{\sigma}\left[\frac{1}{2}\sigma^{ab}\nabu_{(a}a_{b)}-\frac{1}{2}\sigma^{ab}\sigma_{af}\sigma^f_{\phantom{f}b}-\frac{2}{3}\theta\sigma^2
-\frac{1}{2}\sigma^{ab}\left(E_{ab}-\frac{1}{2}\pi_{ab}\right)\right].
\ee
From the definition of $\Sigma_a$, and the commutation rules for the spatial and time derivatives, one gets that $\dot{\Sigma}_a = 
\frac{1}{3}\theta\Sigma_a + \ell\nabu_a\dot{\sigma}-\ell h^b_{\phantom{b}a}(\sigma_{;c}u^c)_{;b}+\ell  
h^d_{\phantom{d}a}(h^b_{\phantom{b}d}\sigma_{;b})_{;c}u^c$. Inserting the result (\ref{sigmadot}) and linearizing then gives
\bea \label{s_evol}
\dot{\Sigma}_a & + & \frac{1}{\sigma^2}\left[ 
\frac{1}{2}\sigma^{bc}\sigma_{bf}\sigma^f_{\phantom{f}c}+\frac{2}{3}\theta\sigma^2+\frac{1}{2}\sigma^{bc}\left(E_{bc}-\frac{1}{2}\pi_{bc}\right)\right]\Sigma_a
\\ \nonumber & + & 
\frac{1}{\sigma}\nabu_a\left[\frac{1}{2}\sigma^{cd}\nabu_{(c}a_{d)}-\frac{1}{2}\sigma^{cd}\sigma_{cf}\sigma^f_{\phantom{f}d}-\frac{2}{3}\theta\sigma^2
-\frac{1}{2}\sigma^{cd}\left(E_{cd}-\frac{1}{2}\pi_{cd}\right)\right] \\ \nonumber
& + & \sigma^c_{\phantom{c}a}\Sigma_c+\ell\theta\sigma a_a = 0.
\eea  
However, it is also possible to solve the shear away without evolving it in time but relating it to the other variables. This is because the time-space 
components of the Einstein field equations yield three additional constraint
equations which we have not exploited this far. These constraints can be put in the 
form
\be
\Sigma_a = \frac{2}{3\ell}Z_a+\Omega_a + \sigma^c_{\phantom{c}a}a_c.
\ee
As a consistency check, one might take the time derivative of this equation and see whether it is compatible with the evolution equation for the shear
we just wrote down in (\ref{s_evol}). It turns out that this constraint is indeed preserved in time. 

We can use this information to eliminate $\Sigma_a$ from the matter evolution equation (\ref{d_evol}). In terms of the quantities appearing
in that equation, we have
\bea \label{sigma}
(1+w)\Sigma_a & = & -\frac{2}{3}\dot{D}_a + \frac{2}{3}\left[\theta + (1+c_s^2+w)\an\right]D_a + (c_s^2-\frac{2}{3})\sigma_a^{\ph{a}b}D_b
\\ \nonumber & + & w\left(\frac{1}{3}\an E_a - \sigma_a^{\ph{a}b}E_b\right) + (1+w)\frac{1}{\rho}\Omega_a 
\\ \nonumber & + & \frac{\ell}{\rho}\left[\frac{2}{3}(\theta+\an)\Pi_a - \sigma_a^{\ph{a}b}\Pi_b+\frac{2}{3}(\an\mathcal{Q}_a-Q_a)
-\sigma_a^{\ph{a}b}\mathcal{Q}_b\right].
\eea
We note that $\Sigma_a$ does of course not vanish even in the general statistically isotropic case.

\subsection{Structure Formation}
\label{structure}

In the case of magnetic fields it has been learned that neglecting the background anisotropy and resorting to an almost-FLRW treatment can 
be a good approximation on large scales when the magnetic field is weak, but the shear may introduce small-scale effects which are accurately captured only by 
the Bianchi I treatment \cite{Tsagas:1999tu}. As we consider a general anisotropic fluid not necessarily reducible to magnetic field plus perfect matter, it is 
thus interesting to study under which conditions we can safely regard the FLRW description as accurate, and under which conditions the background anisotropies 
should be taken into account in order not to miss possible new physical effects.

In the following, we write down the general equations of the last sections in three cases. To close the system of equations, one must specify the anisotropic
properties of the fluid. In subsection (\ref{ccase1}) we assume and study the case that the anisotropy is small. Then, it turns out, the almost-FLRW approach
is valid, with however some direction-dependence of the perturbations. In subsections (\ref{ccase2}) and (\ref{case3}) we do not assume that the anisotropies are 
negligibly small at the background order, but make other simple assumptions about the properties of perturbations to fully specify the system. In these cases,
the shear effects could not be described by the almost-FLRW treatment.
  
\subsubsection{Scalar Equations.}

\label{scalars}

Since the matter density can be described as a scalar field, it is usual to concentrate on the scalar modes of perturbations when studying 
linear evolution in cosmology. To make contact with these studies, it is useful to extract the scalar part of the more general equations in the previous
section. A conventional procedure for this is based on a local decomposition \cite{Ellis:1990gi} of the spatial gradients,
\be
\ell \equiv \nabu^a X_a = \frac{1}{3}h_{ab}X + \left[X_{(ab)}-\frac{1}{3}h_{ab}X\right] + X_{[ab]}.
\ee
As an example, we may act on the density gradient of matter field, $D_a$, with $\nabu^a$, to get the spherically symmetric part of the scalar perturbation.
Explicitly, we define this variable as 
\be \label{def1}
\ell\nabu^aD_a \equiv \Delta.
\ee
In the presence of shear there arise other relevant scalars quantifying the properties of matter distribution. We define them as follows
\bea 
\ell\sigma^{ab}\nabu_aD_b & \equiv & \Delta^{\sigma}, \\
\ell\pi^{ab}\nabu_aD_b & \equiv & \Delta^{\pi}, \\
\ell\sigma^{ac}\sigma_c^{\phantom{c}b}\nabu_aD_b & \equiv & \Delta^{\sigma^2}.
\label{def2}  
\eea
It then follows that
\bea \label{ide1}
\ell \nabu^a \dot{D}_a & = & \dot{\Delta} + \Delta^\sigma, \\
\ell \nabu^a \ddot{D}_a & = & \ddot{\Delta} + 2\dot{\Delta}^\sigma + \theta\Delta^\sigma + \Delta^{\sigma^2} - \Delta^\pi, \\
\ell \nabu^a \sigma_a^{\phantom{a}b}\dot{D}_b & = & \dot{\Delta}^\sigma + \theta\Delta^\sigma + \Delta^{\sigma^2} - \Delta^\pi. 
\label{ide2}
\eea
In a completely analogous way, one may derive scalar variables from other gradients appearing in our considerations. The scalar equations
one then derives by acting on the general equations with $\ell\nabu^a$. It is
straightforward to pick up the new coefficients of the scalar 
variables by using the definitions (\ref{def1}-\ref{def2}) and the identities (\ref{ide1}-\ref{ide2}).

In the most general case, the system is very complicated. In the following, we will comment the physical implications under three
different approximations and the motivations for such approximations: I) The case that the anisotropy is small and can be treated as perturbation,
II) the case that the from of anisotropy perturbation is determined by it's background form III) the case that the the anisotropy is homogeneous 
also at the perturbative level. 

\subsubsection{Case I: Small Anisotropy.}

\label{ccase1}

We make the simplifying assumption that the anisotropy is small. This is justified by the constraints on the 
late-time anisotropies arising from the quadrupole pattern in the CMB and the directional dependence of the 
luminosity-distance. %\cite{Koivisto:2007bp}. 
Thus we regard both the metric $\sigma$ and the fluid $\pi$ as 
 first order quantities in the perturbative equations. Then we are allowed to drop a considerable number of terms from 
the equations, but the largest effects from anisotropy are still taken into account since some first order terms 
remain. Note that would be absent or of a different form in the FLRW case. Thus this seems as a very reasonable starting 
point for the first analysis of the structure formation with an imperfect dark
energy scenario. 

With this assumption the Eq.(\ref{d_evol}) reduces to 
\bea \label{d_evol2}
\ddot{D}_a & + & \left(\frac{2}{3} + c_s^2-2w\right)\theta\dot{D}_a +
\left[(\frac{1}{2}+4w-\frac{3}{2}w^2-3c_s^2)\rho + c_s^2 \nabu^2 \right]D_a 
\\ & = & \nonumber 
- w\nabu^2 E_a  
- 2(1+w)c_s^2\theta\Omega_a + S_a^{(\Pi )}
\eea
where the imperfect fluid terms are given as
\be \label{a_source}
S^{(\Pi )}_a = \frac{\ell}{\rho}\left(\nabu_a\nabu^b\nabu^c\pi_{bc}
+ \theta\dot{\Pi}_a\right) + 3\ell\left(1-w+c_s^2\right)\Pi_a.
\ee
Directly from the metric, the anisotropic terms would contribute only quadratically. From the fluid they contribute 
also at the first order affecting then directly the evolution of the density perturbation as the 
source term (\ref{a_source}). 

By extracting the scalar part of this system as described in subsection (\ref{scalars}), one may
verify that the result is equivalent to the evolution equation for generalized matter when imperfect fluid 
terms are taken into account \cite{Koivisto:2007sq}. One thus arrives at the conclusion that the law governing the  
evolution of overdensities is exactly the same as in an isotropic case when we have neglected the anisotropy terms at the 
background order. However, one should be careful to not deduce that this means that the total evolution is isotropic in such a 
case. The system Eq.(\ref{d_evol2}) presents three equations for the time evolution of three independent spatial gradients of 
the density, and in the general case they could have both different solutions and different forms. In particular, this could be a 
consequence of an intrinsic anisotropic stress source, $\Pi_a$, which could be zero
for a given direction $a$ but not for the others. 
Our formula Eq.(\ref{d_evol2}) thus provides a tool to study fully consistently and covariantly the fundamentally anisotropic 
properties of matter inhomogeneities in cosmology, which may be hidden in the standard description of the overdensity 
scalar perturbation. One may witness the anisotropy by monitoring the solutions of the individual equations of 
the system (\ref{d_evol2}), while summing their gradients yields the scalar quantity which has a straightforward interpretation 
as the overdensity in the standard FLRW cosmology, which at small scales does not depend on the gauge choice.   

This also shows that, even when  a Bianchi universe is close to the FLRW model, one cannot consider the anisotropies in the perturbed Bianchi 
universe as a sum of FLRW perturbations and the small anisotropies of the exact Bianchi model. When an imperfect source is present, its perturbations
(\ref{a_source}) can break explicitly the statistical isotropy already at the linear level. These fluid perturbations of course couple to all other 
perturbations in the metric and in matter. These become then statistically anisotropic as well. Therefore the fluctuations of the model evolve differently
than in a FLRW model, in particular the resulting anisotropy pattern is not a
simple superposition of standard isotropic perturbations and a Bianchi
type perturbation about the FLRW. However, this kind of approximation could be reasonably good when only perfect fluid sources are present. This is 
because the shear in the metric typically enters the equations only trough $\sigma^2$, i.e. quadratically. In such a case, one could expect that some 
effects are given, to first perturbative order, as the sum of ordinary FLRW perturbation and the Bianchi pattern.     

We conclude that when the anisotropy is small, perturbations in each direction behave like in different isotropic universes, each obeying their 
Eq.(\ref{d_evol2}).

\subsubsection{Case II: ''Adiabatic'' Anisotropy.}

\label{ccase2}

Clearly, to perform actual calculations one has to specify the properties of the fluids under consideration. We would still like to keep the discussion general, 
and therefore we apply a parametrization which could approximate a wide range
of different models at a certain limit. We then choose the $w$, $\delta$, 
$\gamma$ parametrization as it seems a natural and straightforward generalization of the concept of equation of state in the FLRW universe. In the Bianchi I 
background, one may fully specify the properties of a fluid with energy density $\rho$ by giving the three pressures in different directions, or equivalently, 
the three equations of state for the fluid. In the following, we assume this holds to the perturbed order. In principle the inhomogeneities in the stresses 
would bring additional degrees of freedom, but our parametrization assumes
they are described by the homogeneous equations of state. This may be interpreted as a 
generalized adiabaticity condition, since in absence of entropy, the pressure perturbation evolution is determined by the homogeneous background evolution.

Recall the decomposition of the fluid content. In Eq.(\ref{perfect_emt1}) we described the energy momentum tensor for the anisotropic component. It may 
be written as 
$$T^\mu_{X\nu} = diag\left(-1,w_X-\delta-\gamma,w_X+2\delta-\gamma,w_X-\delta+2\gamma\right)\hat{\rho}_X.$$ 
%Note that in this section we use a hat for the $\hat{w}$-parameter of the anisotropic component. 
%This is because we do not want to confuse it with the $w$ of the total
%matter content of the universe (which is the previous subsection was treated as a single fluid).     
We are now in a frame having $u^\mu=(1,0,0,0)$. Therefore, the components in the stress-energy tensor (\ref{set}) could now be identified as 
\bea \label{dg_pi}
\rho  & = & \hat{\rho}_X, \qquad p = w_X\rho_X, \qquad q_a = 0, \\
\pi^a_{\ph{a}b} & = & diag\left(0,-\delta-\gamma,2\delta-\gamma,-\delta+2\gamma\right)\rho_X. 
\eea
From the observation that $q_a$ vanishes one could justify to use the equation (\ref{d_evol}) with the heat flux sources $S^{(Q)}$ turned to zero.
If the anisotropy is to be small, the condition is clearly that $\gamma$ and $\delta$ are small. 
One finds that the gradients of the anisotropic stresses are now 
\be \label{pii}
\Pi_a  =  \frac{\rho}{\ell}\left[0,-(\delta+\gamma)D_x,(2\delta-\gamma)D_y,(-\delta+2\gamma)D_z\right],
\ee
and the sum term appearing in the source term Eq.(\ref{s_pi}) is then 
\bea
\frac{\ell}{\rho}\nabu_a\nabu^b\nabu^c\pi_{bc}  & = &  -(\delta+\gamma)\nabu_a\nabu^x D_x \\ \nonumber &+& 
(2\delta-\gamma)\nabu_a\nabu^yD_y - (\delta-2\gamma)\nabu_a\nabu^zD_z.
\eea
The latter source term thus couples to $D_b$-terms with $a \neq b$ into the evolution equation for $D_a$, (\ref{d_evol}).
%The metric shear can now shown to be 
%\be \label{dg_sigma}
%\sigma_{ab} = \frac{H}{3} diag\left(0,R+S,-2R+S,R-2S\right),
%\ee
%which then gives 
%\be \label{sigmai}
%\sigma^2 = \frac{1}{3}\left(R^2+S^2-RS\right)H^2.
%\ee
Since the metric shear $\sigma^2 = \frac{1}{27}\left(R^2+S^2-RS\right)\theta^2$ clearly the small anisotropy limit corresponds to $R,S \ll 1$. 

Let us be more precise about this limit.
One observes that if the coefficients $\delta$ and $\gamma$ are supposed to be perturbatively small, {\it all} contributions to the perturbations equations at 
the linear order vanish. This is because we consider that the constant coefficient of the homogeneous anisotropy gives the relation between the density 
gradient and the (anisotropic) pressure gradient as well as the
relation between the background density and the homogeneous (anisotropic)
pressure. Since the density gradient is already a 
perturbation, the anisotropic pressure gradients (\ref{dg_pi}) should be of the second order in perturbations. Similarly with the metric shear: if $S$ and $R$ are regarded as first order, then all the terms involving $\sigma$ in the evolution equations are higher than first order.  
Therefore the anisotropic terms would decouple from all perturbations equations and the effect of anisotropy would indeed be negligible. Note that the 
homogeneous anisotropy and the gradient of the anisotropy {\it could} be both of the linear order, though it is not the case with the present parametrization. 
Now the simplest nontrivial corrections to the standard FLRW treatment appear if one considers {\it the squares} of the anisotropy variables as first order 
perturbations. 

In general, if the anisotropy terms satisfy such conditions, the system of three equations for the gradients of the matter aggregations can be written as 
\bea \label{d_evol3}
\ddot{D}_a & + & \left(\frac{2}{3}+c_s^2-2w\right)\dot{D}_a + 2\sigma_a^{\ph{a}b}\dot{D}_b
 -  \left(\frac{1}{2}+4w-\frac{3}{2}w^2-3c_s^2\right)\rho D_a
\\ \nonumber
&-& \left(\frac{1}{3}+2w-c_s^2\right)\theta\sigma_a^{\ph{a}b}D_b
-  \nabu^2 \left(c_s^2D_a + w E_a\right)
+ 2(1+w)c_s^2\theta\Omega_a
\\ \nonumber
&=&   
4\sigma(1+w)\Sigma_a +
\frac{\ell}{\rho}\Big[\nabu_a\nabu^b\nabu^c\pi_{bc} - \dot{\Xi}_a - \left(\frac{4}{3} + w\right)\theta\Xi_a 
\\ \nonumber 
& + & \theta\dot{\Pi}_a
+ 3\left(1-w+c_s^2\right)\rho\Pi_a \Big].
\nonumber
\eea
Now, in addition to the imperfect source stresses, the coupling of the metric shear will generate anisotropies in the perturbations. 

We conclude that there exists a generalization of the adiabatic condition to Bianchi I universes which then uniquely determine the perturbative 
stresses of the fluid.
 
\subsubsection{Case III: Smooth Anisotropy.}

\label{case3}

Yet one more physically motivated approximation is to consider the anisotropy field as smooth. A dark energy -like component in usual cases is 
considered to be rather smooth, especially at small scales. In fact no evidence for clumping of dark energy has been found. Slowly rolling light 
fields with minimal couplings tend to have negligible perturbations. For the cosmological constant this becomes an exact covariant statement: there 
are no cosmological fluctuations in the usual $\Lambda$-term. 
When the pressure varies in direction and with time, its spatial gradients may be only approximately 
zero. Basically, one then considers the effects of the anisotropic background expansion on the matter inhomogeneities. To do this consistently one 
should couple not only the metric $\sigma$ but also the fluid $\pi$ anisotropy to the equations. Then we can, however, neglect the possible entropy in the 
perturbations by a direct analogy with the usual $\Lambda$ term. One notes that the entropy couples also to the large scale perturbations 
through the anisotropic terms, whereas in an FLRW universe one may neglect $E$ in the limit $\nabu^2 \rightarrow 0$. 

Indeed, for the model of subsection (\ref{ccase1}), one could consider a very tidy set of  
perturbation equations where the isotropic perturbations of the $\Lambda$-term identically vanish, and the condition 
$\an = 0$ allows to drop a large number of terms. For the variant of the model, discussed in subsection \ref{case2}, the isotropic pressure of the 
$\Lambda$-term is not constant and therefore cannot be smooth. This introduces entropy between the $\Lambda$-term and matter. 
One would then consider the evolution of the total density gradient $D_a$, as that is simply related to the matter density gradient $D^{(m)}_a$ by 
$D^{(m)}_a = (1-U)D_a$. The anisotropic stress and the metric shear at the background level would then be 
\be
\pi^{a}_{\phantom{a}b} = diag\left[0,-1,2,-1\right]\delta\rho_\Lambda
\ee 
\be
\sigma^{a}_{\phantom{a}b} = \frac{1}{3}diag\left[0,1,-2,1\right] HR,
\ee 
giving then $\sigma^2=(HR)^2/3$. Note that $\pi^a_b$ is constant (when $\delta$ is), and thus its fluctuations $\Pi_a$ vanish.
However, similar conclusion does not hold for the $\sigma$-terms: the metric shear has fluctuations and generally $\Sigma_a$ 
is nonzero. Therefore in the model with the constant anisotropic pressure field generates effects on the first order fluctuations. 
In fact this is already reflected in the fact that the total sound speed squared of the system (\ref{ccsound}) can become nonzero when $\delta$ is nonzero.
Because there is entropy present, the possible negativity of the sound speed does not necessarily lead to instabilities.  
These effects could be used to constrain further the scenario proposed in subsection \ref{case2}.

We conclude that either the isotropic or the anisotropic pressure of an imperfect fluid always fluctuates.

\section{Conclusions and Discussion}
\label{conclusions}

We investigated the possibility that the accelerated expansion of our universe, both during the early 
inflation and the 
present dark energy domination, is driven by a component with an anisotropic equation of state. The motivation for this investigation
comes from the frequent appearance vector fields and possible Lorentz violations in fundamental theories, from the need to the test the robustness of 
the basic assumptions of cosmology, and from the hints of statistical anisotropy of our universe that several observations seem to suggest. 

From the examination of the background dynamics we learned that

\begin{itemize}

\item There exist (anisotropic) scaling solutions even without coupling between the components.

\item An anisotropic generalization of the $\Lambda$-term has nontrivial dynamics given if its isotropic pressure is time-varying.

\item The quadropole constraint on the anisotropy may be avoided, and the present SNIa data allows large anisotropy.  

\end{itemize}

From the analysis of system of perturbations the following was found

\begin{itemize}

\item Even in the limit of weak anisotropy, the linearized perturbations will be direction-dependent.

\item The linearized anisotropy is the sum of FLRW perturbations and Bianchi I background only for perfect-fluid cosmology.

\item Either in the limit of vanishing entropy or in the limit of homogeneous field the almost FLRW treatment is not valid. 

\end{itemize}

Let us elaborate on these and other results.

We started by presenting a phenomenological description of a such an anisotropy in terms of the skewness parameters $\delta$ and
$\gamma$, and derived the basic equations describing the model as a dynamical system, and obtained the generic asymptotic
evolution of the universe in some simple cases. Several scaling solutions were found both within the matter dominated epoch as well 
as during an accelerated expansion phase (with the summary of fixed points presented in the Table \ref{tab1}). This seems to be 
interesting, since while in the FLRW universe it has been proven difficult to address the coincidence problem by finding a model
entering from a matter-dominated scaling solution to an accelerating scaling solution, allowing for the presence of three expansion rates 
now opens up the possibility of describing a universe entering from a perfect fluid dominated scaling to an anisotropically accelerating 
scaling era, which might eventually help to understand the coincidence
problem. 
%We show a generalization of that in the more general setting taking into account vector fields. 
In that framework, it is easier to find models with natural parameter values avoiding the fine-tuning problems of cosmological constant or scalar field
quintessence \cite{Boehmer:2007qa,Koivisto:2007bp,Jimenez:2008au}.  

We also contemplated upon the possibility of an anisotropic 
generalization of the cosmological constant constant term. Interestingly, we
found that a model with a constant density and constant anisotropic 
pressures ($\rho = \rho_0$, $\delta = \delta_0$, $\gamma = \gamma_0$) is indeed a consistent covariant modification of the Einstein's cosmological term 
which could generate a viable expansion history (with small enough $\delta_0$ and $\gamma_0$) that exhibits nontrivial anisotropic properties. 
We mentioned noncommutative properties of spacetime as a possible origin of anisotropies in a cosmological $\Lambda$-term. 
One could consider different motivations by writing the vacuum energy as an integral of the zero-point energies over the $\bf{k}$ up to some cut-off 
(summed over every possible field).
If there were any isotropy-breaking physical principle which allowed only wavevectors 
with specific directions, the vacuum energy would acquire nonzero $\delta$ and $\gamma$. More specifically, if the cut-off depended not only on the magnitude
$k$ of the wavevector ${\bf{k}}$, but also on its direction (i.e. some directions would be 
excluded from the integral or contribute with less weight to the vacuum energy) this could explain both the
smallness of the observed cosmological constant and the apparent anisotropic 
features in the cosmological data. 
  
The amount of possible anisotropy, as described by the skewness parameters, can be constrained by the CMB and SNIa 
observations. We found that the present CMB data constraints can be much tighter than the constraints from the anisotropies in 
the luminosity-distance -redshift relationship of SNIa. However, there are classes of models where the CMB bounds from the background 
quadrupole anisotropy can be avoided altogether while there is significant anisotropy at late times that could be observed
in the late SNIa distributions and which could leave statistically anisotropic imprints at the large-angle CMB.
We point out, however, that in more general models, e.g. with time-varying $\delta$ and $\gamma$,
one could allow more anisotropy. It is in principle possible for an arbitrarily anisotropic expansion to escape
detection from CMB (considering only the effects from the background), as long as the
expansion rates evolve in such a way that $e_z=e_y=0$. In other words, the (background)
quadrupole vanishes, if each scale factor has expanded - no matter how anisotropically -
the same amount since the last scattering. 
%This may be demonstrated explicitly with a vector
%field theory example. This is a fine-tuned model, but proves that even very large anisotropies could exist in the expansion 
%history, however, with the quadrupole amplitude, Eqs.(\ref{multipole}, \ref{quad}), nearly or completely unchanged. 

We also considered the inhomogeneous cosmology of these models. The covariant formalism was employed to derive the general evolution equations for the 
perturbations in a universe with background anisotropies, i.e. imperfect fluid
terms and shear of the metric. 
Our formalism and equations thus provide a tool to study fully consistently and covariantly the fundamentally anisotropic 
properties of matter inhomogeneities in cosmology, which may be hidden in the standard description of the overdensity 
scalar perturbation. One may witness the anisotropy by monitoring the solutions of the individual equations of 
the system (\ref{d_evol2}), while summing their gradients yields the scalar quantity which has a straightforward interpretation 
as the overdensity in the standard FLRW cosmology, which at small scales does
not depend on the gauge choice.

The implications of anisotropic sources for the formation of structures 
were discussed in several limiting cases. In particular, in the cases  examined, the deviations from isotropy are small enough that one may treat
all the imperfect terms as first order perturbations. We noticed that this gives a result which is not obtainable by a straightforward generalization of the usual 
FLRW equations. If one considers the evolution equation of matter perturbations (in its closed form (\ref{d_evol2})), one notices that while the metric shear 
would enter only at the second order, the imperfect fluid terms are present at the first order level. Thus, even when the Bianchi universe would be close to the 
FLRW model, one cannot consider the anisotropies in the perturbed Bianchi universe as a sum of FLRW perturbations and the small anisotropies of the exact Bianchi 
model. This is seen from the fact that when an imperfect source is present, its perturbations (\ref{a_source}) can break explicitly the statistical 
isotropy already at the linear level. In general, the perturbation structure of these models allows a rich array of possibilities. The anisotropic properties seem 
to usually extend to the smaller scales also, but this feature is model-dependent and the constraints it implies must be studied on a case-by-case basis.

The present SNIa data allows anisotropic acceleration, but future SNAP data could set things straight about the skewness of dark energy 
and so of its nature. Such possibility would open a completely new window not only on the nature of the CMB anomalies but
also into high energy physics models beyond the usual isotropic candidates of
inflation or dark energy models such as the inflaton, the quintessence 
scalar fields or even an isotropic cosmological constant. Hence, an anisotropic equation of state may not only successfully unify the early inflation with 
the present days acceleration, but might also be the culprit for both the apparent problem of cosmic coincidence as well as the large-angle anomalies in 
the CMB.

The 
detailed signatures of the anisotropies, either originating from the early or the late accelerating epoch of the universe, to the full 
sky maps of the CMB should be determined: could these signatures rule some or all of the models out, or could they in fact be identified 
with the observed anomalies? Investigations along these lines are ongoing.

\section{Appendix: Pedestrian notation for Bianchi I}

In this Appendix we review our system within the coordinate formalism. Our main developments in the bulk of the paper follow the covariant approach
for the sake of homogeneity. However, for some considerations the full covariant machinery is unnecessary, and analysis is simpler and perhaps
more transparent by specifying the coordinate system and writing the equations in terms of the metric components. 

The Bianchi I metric as generalization of the flat 
FLRW metric is
\be \label{metric}
ds^2 = -dt^2 +a^2(t)dx^2+b^2(t)dy^2 + c^2(t)dz^2.
\ee
Due to the presence of three scale factors $a$, $b$ and $c$,  one has to introduce also three expansion rates. 
In principle all these could be different, and in the limiting case that all of them are equal, one of course recovers the FLRW case.
Note that we use the roman letters to label the covariant indices, like in the expression for the energy momentum 
tensor (\ref{set}), whereas we use the Greek letters to label the indices of tensors when we refer to the components in the specific choice of basis
(\ref{metric}). 
The perfect fluid energy momentum tensor is then simply
\be \label{perfect_emt}
T^\mu_{(m)\nu} = diag(-1,w_m,w_m,w_m)\rho_m,
\ee
In the coordinate system we have chosen, the energy-momentum tensor must be diagonal, since the nondiagonal Einstein tensor
vanishes identically. However, the pressure in each spatial direction could be different. We then parametrize
the anisotropy of the pressure by the two skewness parameters
\be
\label{paramsa}
3\delta \equiv (p_y-p_x)/\rho, \quad 3\gamma \equiv (p_z-p_x)/\rho.
\ee
The
energy-momentum tensor of this fluid is then
\be \label{imperfect_emt}
T^\mu_{\phantom{\mu}\nu} = diag(-1,w,w+3\delta,w + 3\gamma)\rho.
\ee
The Einstein equations would not allow non-diagonal components for the stresses with the metric (\ref{metric}).

With the matter content (\ref{perfect_emt}) and (\ref{imperfect_emt}), the three nontrivial spatial components
of the field equations $$G^\mu_{\phantom{\mu}\nu} = 8\pi G (T^\mu_{(m)\nu}+T^\mu_{\phantom{\mu}\nu})$$ are given as 
\be \label{fieldxx}
\frac{\ddot{b}}{b} +
\frac{\ddot{c}}{c} +
\frac{\dot{b}\dot{c}}{bc} = -8\pi G
\left[w_m \rho_m  + w\rho  \right],
\ee
\be \label{fieldyy}
\frac{\ddot{a}}{a} +
\frac{\ddot{c}}{c} +
\frac{\dot{a}\dot{c}}{ac} = -8\pi G
\left[w_m \rho_m  + \left(w+3\delta\right) \rho \right],
\ee
\be \label{fieldzz}
\frac{\ddot{a}}{a} +
\frac{\ddot{b}}{b} +
\frac{\dot{a}\dot{b}}{ab} = -8\pi G
\left[w_m \rho_m  + \left(w+3\gamma\right) \rho \right],
\ee
and the time component is
\be \label{fieldtt}
\frac{\dot{a}\dot{b}}{ab} +
\frac{\dot{b}\dot{c}}{bc} +
\frac{\dot{c}\dot{a}}{ca} = 8\pi G \left[\rho_m + \rho\right].
\ee
Overdots denote derivatives with respect to time $t$.

The continuity equations are then given by the divergence of the energy-momentum tensors.
We let the two components also interact, and thus allow nonzero divergence for the individual components.
\be \label{cont1a}
\dot{\rho}_m + 3H\left(1+w_m\right)\rho_m = Q H \rho,
\ee
and
\be \label{cont2a}
\dot{\rho}   + H\left[3\left(1+w\right) + \delta\left(3 -2R +S\right) + \gamma\left(3+R-2S\right) \right]\rho = - Q H \rho,
%+ 3\delta \frac{\dot{b}}{b}\rho + 3\frac{\dot{c}}{c}\rho = -Q H \rho,
\ee
where $Q$ describes the coupling. If there are no interactions, one finds that the densities scale as
$$\rho_m \sim (abc)^{-1-w_m}, \qquad \rho \sim (abc)^{-1-w}b^{3\delta}c^{3\gamma}.$$ The time variable we will
use is an average e-folding time $x$, defined as follows:
\be \label{efolding}
x \equiv \frac{1}{3}\log{(abc)}.
\ee
One notes that $H = \dot{x}$, in analogy with the FLRW case, where the e-folding time reduces to $x=\log{a}$.
Derivative of a function $f$ with respect to $x$ will be denoted by prime, unless $f$ is explicitly specified as a function $f(y)$ of some other variable $y$.

In the following we will rewrite this system in a more convenient form for analytical and numerical
study. 
It is then useful to follow the notation introduced in Ref. \cite{Barrow:1997sy} by
expressing the mean expansion rate as an average Hubble rate $H$
\be
H \equiv \frac{1}{3}\left(\frac{\dot{a}}{a}+\frac{\dot{b}}{b} + \frac{\dot{c}}{c}\right),
\ee
and the differences of the expansion rates as the Hubble-normalized shear $R$ and $S$
\be \label{shear}
R  \equiv \frac{1}{H}\left(\frac{\dot{a}}{a}-\frac{\dot{b}}{b}\right), \quad
S  \equiv \frac{1}{H}\left(\frac{\dot{a}}{a}-\frac{\dot{c}}{c}\right).
\ee
Assuming that all the expansion rates are positive, one has always $-3 < R,S < 3$. This is true also if either $R=0$ or $S=0$.
We find then a generalized Friedmann equation of the form
\be \label{friedmanna}
H^2 = \frac{8\pi G}{3}\frac{\rho_m+\rho}{1-\frac{1}{9}\left(R^2+ S^2 - RS\right)}.
\ee

%In general one may have 
%\be
%-2\sqrt{3}<R<2\sqrt{3},   
%\ee
%when
%\be
%\frac{R}{2}-\frac{\sqrt{3}}{2}\sqrt{12-R^2} \ge S \ge 
%\frac{R}{2}+\frac{\sqrt{3}}{2}\sqrt{12-R^2}. 
%\ee

Our dynamical variables are then the density fraction $U$, and the two shear anisotropies $R$ and
$S$. The evolution equations for these are then
\be \label{sys1a}
  U' =
       U\left(U -1\right)
            \left[\gamma \left(3+R-2S \right)  +
              \delta\left(3-2R+S\right) + 3\left(w - w_m\right) \right] - UQ,
\ee
\be \label{s_eqa}
  S' =
       \frac{1}{6}
       \left(9 - R^2 + RS - S^2 \right)
       \left\{ S\left[U\left(\delta + \gamma+w-w_m\right)+w_m-1\right]-6\gamma U \right\},
\ee
\be \label{r_eqa}
  R' =
       \frac{1}{6}
       \left(9 - R^2 + RS - S^2 \right)
       \left\{ R\left[U\left(\delta + \gamma+w-w_m\right)+w_m-1\right]-6\delta U \right\}.
\ee

%%%%%%%%%%%%%%%%%%%%%%%%%%%%%%%%%%%%%%%%%%%%%%%%%%%%%%%%%%%%%%%%%%%%%%
\section*{Acknowledgments} %%%%%%%%%%%%%%%%%%%%%%%%%%%%%%%%%%%%%%%%%%%
%%%%%%%%%%%%%%%%%%%%%%%%%%%%%%%%%%%%%%%%%%%%%%%%%%%%%%%%%%%%%%%%%%%%%%

DFM acknowledges support from the A. Humboldt Foundation. We thank the anonymous referee for 
critical reading of the manuscript and several suggestions which have helped us to improve the paper.

%%%%%%%%%%%%%%%%%%%%%%%%%%%%%%%%%%%%%%%%%%%%%%%%%%%%%%%%%%%%%%%%%%%%%%
\section*{References} %%%%%%%%%%%%%%%%%%%%%%%%%%%%%%%%%%%%%%%%%%%%%%%%
%%%%%%%%%%%%%%%%%%%%%%%%%%%%%%%%%%%%%%%%%%%%%%%%%%%%%%%%%%%%%%%%%%%%%%

\bibliography{birefs1}

\begin{thebibliography}{10}

\bibitem{Eriksen:2003db}
H.~K. Eriksen, F.~K. Hansen, A.~J. Banday, K.~M. Gorski, and P.~B. Lilje.
\newblock Asymmetries in the cmb anisotropy field.
\newblock {\em Astrophys. J.}, 605:14--20, 2004.

\bibitem{Hinshaw:2006ia}
G.~Hinshaw et~al.
\newblock Three-year wilkinson microwave anisotropy probe (wmap) observations:
  Temperature analysis.
\newblock {\em astro-ph/0603451}, 2006.

\bibitem{Land:2005ad}
Kate Land and Joao Magueijo.
\newblock The axis of evil.
\newblock {\em Phys. Rev. Lett.}, 95:071301, 2005.

\bibitem{Rakic:2007ve}
Aleksandar Rakic and Dominik~J. Schwarz.
\newblock Correlating anomalies of the microwave sky: The good, the evil and
  the axis.
\newblock {\em astro-ph/0703266}, 2007.

\bibitem{Magueijo:2006we}
Joao Magueijo and Rafael~D. Sorkin.
\newblock Occam's razor meets wmap.
\newblock {\em Mon. Not. Roy. Astron. Soc. Lett.}, 377:L39--L43, 2007.

\bibitem{Land:2006bn}
Kate Land and Joao Magueijo.
\newblock The axis of evil revisited.
\newblock {\em astro-ph/0611518}, 2006.

\bibitem{Copi:2005ff}
Craig~J. Copi, D.~Huterer, D.~J. Schwarz, and G.~D. Starkman.
\newblock On the large-angle anomalies of the microwave sky.
\newblock {\em Mon. Not. Roy. Astron. Soc.}, 367:79--102, 2006.

\bibitem{Hajian:2006ud}
Amir Hajian and Tarun Souradeep.
\newblock Testing global isotropy of three-year wilkinson microwave anisotropy
  probe (wmap) data: Temperature analysis.
\newblock {\em Phys. Rev.}, D74:123521, 2006.

\bibitem{Gordon:2005ai}
Christopher Gordon, Wayne Hu, Dragan Huterer, and Tom Crawford.
\newblock Spontaneous isotropy breaking: A mechanism for cmb multipole
  alignments.
\newblock {\em Phys. Rev.}, D72:103002, 2005.

\bibitem{Armendariz-Picon:2005jh}
Cristian Armendariz-Picon.
\newblock Footprints of statistical anisotropies.
\newblock {\em JCAP}, 0603:002, 2006.

\bibitem{Gumrukcuoglu:2006xj}
A.~E. Gumrukcuoglu, Carlo~R. Contaldi, and Marco Peloso.
\newblock Cmb anomalies from relic anisotropy.
\newblock {\em astro-ph/0608405}, 2006.

\bibitem{Ackerman:2007nb}
Lotty Ackerman, Sean~M. Carroll, and Mark~B. Wise.
\newblock Imprints of a primordial preferred direction on the microwave
  background.
\newblock {\em Phys. Rev.}, D75:083502, 2007.

\bibitem{Alexander:2006mt}
Stephon H.~S. Alexander.
\newblock Is cosmic parity violation responsible for the anomalies in the wmap
  data?
\newblock {\em hep-th/0601034}, 2006.

\bibitem{Armendariz-Picon:2007nr}
C.~Armendariz-Picon.
\newblock Creating statistically anisotropic and inhomogeneous perturbations.
\newblock {\em arXiv:0705.1167 [astro-ph]}, 2007.

\bibitem{Bohmer:2007ut}
Christian~G. Bohmer and David~Fonseca Mota.
\newblock Cmb anisotropies and inflation from non-standard spinors.
\newblock {\em arXiv:0710.2003 [astro-ph]}, 2007.

\bibitem{deOliveira-Costa:2006zj}
Angelica de~Oliveira-Costa and Max Tegmark.
\newblock Cmb multipole measurements in the presence of foregrounds.
\newblock {\em Phys. Rev.}, D74:023005, 2006.

\bibitem{Inoue:2006rd}
Kaiki~Taro Inoue and Joseph Silk.
\newblock Local voids as the origin of large-angle cosmic microwave background
  anomalies.
\newblock {\em Astrophys. J.}, 648:23--30, 2006.

\bibitem{Enqvist:2006cg}
Kari Enqvist and Teppo Mattsson.
\newblock The effect of inhomogeneous expansion on the supernova observations.
\newblock {\em JCAP}, 0702:019, 2007.

\bibitem{Armendariz-Picon:2004pm}
Christian Armendariz-Picon.
\newblock Could dark energy be vector-like?
\newblock {\em JCAP}, 0407:007, 2004.

\bibitem{Boehmer:2007qa}
C.~G. Boehmer and T.~Harko.
\newblock Dark energy as a massive vector field.
\newblock {\em Eur. Phys. J.}, C50:423--429, 2007.

\bibitem{Libanov:2007mq}
Maxim Libanov, Valery Rubakov, Eleftherios Papantonopoulos, M.~Sami, and Shinji
  Tsujikawa.
\newblock Uv stable, lorentz-violating dark energy with transient phantom era.
\newblock {\em JCAP}, 0708:010, 2007.

\bibitem{us}
Tomi~S. Koivisto and David~F. Mota.
\newblock Vector field models of inflation and dark energy.
\newblock {\em 0805.4229}, 2008.

\bibitem{Koivisto:2005mm}
Tomi Koivisto and David~F. Mota.
\newblock Dark energy anisotropic stress and large scale structure formation.
\newblock {\em Phys. Rev.}, D73:083502, 2006.

\bibitem{Battye:2006mb}
Richard~A. Battye and Adam Moss.
\newblock Anisotropic perturbations due to dark energy.
\newblock {\em Phys. Rev.}, D74:041301, 2006.

\bibitem{o1}
P.~R. Pereira, M.~F.~A. Da~Silva, and R.~Chan.
\newblock Anisotropic self-similar cosmological model with dark energy.
\newblock {\em Int. J. Mod. Phys.}, D15:991--999, 2006.

\bibitem{o2}
Kishore~N. Ananda and Marco Bruni.
\newblock Cosmo-dynamics and dark energy with a quadratic eos: Anisotropic
  models, large-scale perturbations and cosmological singularities.
\newblock {\em Phys. Rev.}, D74:023524, 2006.

\bibitem{o3}
Bijan Saha.
\newblock Anisotropic cosmological models with perfect fluid and dark energy
  revisited.
\newblock {\em Int. J. Theor. Phys.}, 45:952--964, 2006.

\bibitem{o4}
Bijan Saha.
\newblock Anisotropic cosmological models with perfect fluid and dark energy.
\newblock {\em Chin. J. Phys.}, 43:1035--1043, 2005.

\bibitem{val1}
Valeria Pettorino, C.~Baccigalupi, and F.~Perrotta.
\newblock Scaling solutions in scalar-tensor cosmologies.
\newblock {\em JCAP}, 0512:003, 2005.

\bibitem{val2}
Valeria Pettorino, Carlo Baccigalupi, and Gianpiero Mangano.
\newblock Extended quintessence with an exponential coupling.
\newblock {\em JCAP}, 0501:014, 2005.

\bibitem{Koivisto:2006ie}
Tomi Koivisto.
\newblock {The matter power spectrum in $f(R)$ gravity}.
\newblock {\em Phys. Rev.}, D73:083517, 2006.

\bibitem{Koivisto:2005yc}
Tomi Koivisto and Hannu Kurki-Suonio.
\newblock Cosmological perturbations in the palatini formulation of modified
  gravity.
\newblock {\em Class. Quant. Grav.}, 23:2355--2369, 2006.

\bibitem{clifton}
Timothy Clifton, David~F. Mota, and John~D. Barrow.
\newblock Inhomogeneous gravity.
\newblock {\em Mon. Not. Roy. Astron. Soc.}, 358:601, 2005.

\bibitem{skordis}
Constantinos Skordis, D.~F. Mota, P.~G. Ferreira, and C.~Boehm.
\newblock Large scale structure in bekenstein's theory of relativistic mond.
\newblock {\em Phys. Rev. Lett.}, 96:011301, 2006.

\bibitem{amar}
Morad Amarzguioui, O.~Elgaroy, D.~F. Mota, and T.~Multamaki.
\newblock {Cosmological constraints on $f(R)$ gravity theories within the
  Palatini approach}.
\newblock {\em Astron. Astrophys.}, 454:707--714, 2006.

\bibitem{doug1}
David~F. Mota and Douglas~J. Shaw.
\newblock Strongly coupled chameleon fields: New horizons in scalar field
  theory.
\newblock {\em Phys. Rev. Lett.}, 97:151102, 2006.

\bibitem{doug2}
David~F. Mota and Douglas~J. Shaw.
\newblock Evading equivalence principle violations, astrophysical and
  cosmological constraints in scalar field theories with a strong coupling to
  matter.
\newblock {\em Phys. Rev.}, D75:063501, 2007.

\bibitem{Koivisto:2005nr}
Tomi Koivisto.
\newblock Growth of perturbations in dark matter coupled with quintessence.
\newblock {\em Phys. Rev.}, D72:043516, 2005.

\bibitem{luca}
Luca Amendola.
\newblock Coupled quintessence.
\newblock {\em Phys. Rev.}, D62:043511, 2000.

\bibitem{Caldwell:2007cw}
Robert Caldwell, Asantha Cooray, and Alessandro Melchiorri.
\newblock Constraints on a new post-general relativity cosmological parameter.
\newblock {\em astro-ph/0703375}, 2007.

\bibitem{Ichiki:2007vn}
Kiyotomo Ichiki and Tomo Takahashi.
\newblock Constraints on generalized dark energy from recent observations.
\newblock {\em astro-ph/0703549}, 2007.

\bibitem{Mota:2007sz}
D.~F. Mota, J.~R. Kristiansen, T.~Koivisto, and N.~E. Groeneboom.
\newblock Constraining dark energy anisotropic stress.
\newblock {\em arXiv:0708.0830 [astro-ph]}, 2007.

\bibitem{brook1}
Anthony~W. Brookfield, C.~van~de Bruck, D.~F. Mota, and D.~Tocchini-Valentini.
\newblock {Cosmology of mass-varying neutrinos driven by quintessence: Theory
  and observations}.
\newblock {\em Phys. Rev.}, D73:083515, 2006.

\bibitem{amendola}
Luca Amendola, Martin Kunz, and Domenico Sapone.
\newblock Measuring the dark side (with weak lensing).
\newblock {\em arXiv:0704.2421 [astro-ph]}, 2007.

\bibitem{uzan}
Carlo Schimd, Jean-Philippe Uzan, and Alain Riazuelo.
\newblock Weak lensing in scalar-tensor theories of gravity.
\newblock {\em Phys. Rev.}, D71:083512, 2005.

\bibitem{bruck}
D.~F. Mota and C.~van~de Bruck.
\newblock {On the spherical collapse model in dark energy cosmologies}.
\newblock {\em Astron. Astrophys.}, 421:71--81, 2004.

\bibitem{nunes}
Nelson~J. Nunes and D.~F. Mota.
\newblock {Structure Formation in Inhomogeneous Dark Energy Models}.
\newblock {\em Mon. Not. Roy. Astron. Soc.}, 368:751--758, 2006.

\bibitem{Koivisto:2007bp}
Tomi Koivisto and David~F. Mota.
\newblock Accelerating cosmologies with an anisotropic equation of state.
\newblock {\em arXiv:0707.0279 [astro-ph]}, 2007.

\bibitem{Barrow:1997sy}
John~D. Barrow.
\newblock Cosmological limits on slightly skew stresses.
\newblock {\em Phys. Rev.}, D55:7451--7460, 1997.

\bibitem{Barrow:2005df}
John~D. Barrow, Yoshida Jin, and Kei-ichi Maeda.
\newblock Cosmological co-evolution of yang-mills fields and perfect fluids.
\newblock {\em Phys. Rev.}, D72:103512, 2005.

\bibitem{Campanelli:2006vb}
Luigi Campanelli, P.~Cea, and L.~Tedesco.
\newblock Ellipsoidal universe can solve the cmb quadrupole problem.
\newblock {\em Phys. Rev. Lett.}, 97:131302, 2006.

\bibitem{BeltranJimenez:2007ai}
Jose Beltran~Jimenez and Antonio~L. Maroto.
\newblock Cosmology with moving dark energy and the cmb quadrupole.
\newblock {\em astro-ph/0703483}, 2007.

\bibitem{Rodrigues:2007ny}
Davi~C. Rodrigues.
\newblock Anisotropic cosmological constant and the cmb quadrupole anomaly.
\newblock {\em arXiv:0708.1168 [astro-ph]}, 2007.

\bibitem{Longo:2007gr}
Michael~J. Longo.
\newblock Does the universe have a handedness.
\newblock {\em astro-ph/0703325}, 2007.

\bibitem{Longo:2007pc}
Michael~J. Longo.
\newblock Is the cosmic axis of evil due to a large-scale magnetic field.
\newblock {\em astro-ph/0703694}, 2007.

\bibitem{barrow1}
John~D. Barrow and Sigbjorn Hervik.
\newblock Anisotropically inflating universes.
\newblock {\em Phys. Rev.}, D73:023007, 2006.

\bibitem{barrow2}
John~D. Barrow and Sigbjorn Hervik.
\newblock On the evolution of universes in quadratic theories of gravity.
\newblock {\em Phys. Rev.}, D74:124017, 2006.

\bibitem{burgess}
C.~P. Burgess, Richard Easther, Anupam Mazumdar, David~F. Mota, and Tuomas
  Multamaki.
\newblock Multiple inflation, cosmic string networks and the string landscape.
\newblock {\em JHEP}, 05:067, 2005.

\bibitem{easson}
Damien~A. Easson, Ruth Gregory, David~F. Mota, Gianmassimo Tasinato, and
  I.~Zavala.
\newblock {Spinflation}.
\newblock {\em JCAP}, 0802:010, 2008.

\bibitem{Ellis:1998ct}
George F.~R. Ellis and Henk van Elst.
\newblock Cosmological models.
\newblock {\em gr-qc/9812046}, 1998.

\bibitem{Buniy:2005qm}
Roman~V. Buniy, Arjun Berera, and Thomas~W. Kephart.
\newblock Asymmetric inflation: Exact solutions.
\newblock {\em Phys. Rev.}, D73:063529, 2006.

\bibitem{Saha:2007if}
Bijan Saha.
\newblock Interacting spinor and scalar fields in bianchi type-i universe
  filled with viscous fluid: Exact and numerical solutions.
\newblock {\em gr-qc/0703124}, 2007.

\bibitem{Saha:2004qt}
Bijan Saha.
\newblock Bianchi type i universe with viscous fluid.
\newblock {\em Mod. Phys. Lett.}, A20:2127--2144, 2005.

\bibitem{Brevik:2001ed}
I.~Brevik and S.~D. Odintsov.
\newblock On the cardy-verlinde entropy formula in viscous cosmology.
\newblock {\em Phys. Rev.}, D65:067302, 2002.

\bibitem{Brevik:2005bj}
I.~Brevik and O.~Gorbunova.
\newblock Dark energy and viscous cosmology.
\newblock {\em Gen. Rel. Grav.}, 37:2039--2045, 2005.

\bibitem{Brevik:2006wa}
Iver Brevik.
\newblock Crossing of the w = -1 barrier in two-fluid viscous modified gravity.
\newblock {\em Gen. Rel. Grav.}, 38:1317--1328, 2006.

\bibitem{King:2006cy}
Emma~J. King and Peter Coles.
\newblock Dynamics of a magnetised bianchi i universe with vacuum energy.
\newblock {\em Class. Quant. Grav.}, 24:2061--2072, 2007.

\bibitem{Raj:2007ug}
Bali Raj, Pareek~Umesh Kumar, and Pradhan Anirudh.
\newblock Bianchi type i massive string magnetized barotropic perfect fluid
  cosmological model in general relativity.
\newblock {\em arXiv:0704.0753 [gr-qc]}, 2007.

\bibitem{Pontzen:2007ii}
Andrew Pontzen and Anthony Challinor.
\newblock Bianchi model cmb polarization and its implications for cmb
  anomalies.
\newblock {\em arXiv:0706.2075 [astro-ph]}, 2007.

\bibitem{Ando:2007hc}
Shin'ichiro Ando and Marc Kamionkowski.
\newblock Nonlinear evolution of anisotropic cosmological power.
\newblock {\em arXiv:0711.0779 [astro-ph]}, 2007.

\bibitem{Pitrou:2008gk}
Cyril Pitrou, Thiago~S. Pereira, and Jean-Philippe Uzan.
\newblock {Predictions from an anisotropic inflationary era}.
\newblock {\em arXiv:0801.3596 [astro-ph]}, 2008.

\bibitem{Bonvin:2005ps}
Camille Bonvin, Ruth Durrer, and M.~Alice Gasparini.
\newblock Fluctuations of the luminosity distance.
\newblock {\em Phys. Rev.}, D73:023523, 2006.

\bibitem{Schwarz:2007wf}
Dominik~J. Schwarz and Bastian Weinhorst.
\newblock (an)isotropy of the hubble diagram: comparing hemispheres.
\newblock {\em arXiv:0706.0165 [astro-ph]}, 2007.

\bibitem{Riess:2006fw}
Adam~G. Riess et~al.
\newblock New hubble space telescope discoveries of type ia supernovae at $z >
  1$: Narrowing constraints on the early behavior of dark energy.
\newblock {\em astro-ph/0611572}, 2006.

\bibitem{Astier:2005qq}
Pierre Astier et~al.
\newblock The supernova legacy survey: Measurement of $omega_m$, $omega_lambda$
  and w from the first year data set.
\newblock {\em Astron. Astrophys.}, 447:31--48, 2006.

\bibitem{Perko:1972cs}
T.~E. Perko, R.~A. Matzner, and L.~C. Shepley.
\newblock Galaxy formation in anisotropic cosmologies.
\newblock {\em Phys. Rev.}, D6:969--983, 1972.

\bibitem{Noh:1987vk}
H.~Noh and J.~C. Hwang.
\newblock Perturbations of an anisotropic space-time: Formulation.
\newblock {\em Phys. Rev.}, D52:1970--1987, 1995.

\bibitem{Miedema:1992wb}
P.~G. Miedema and W.~A. van Leeuwen.
\newblock Perturbations in bianchi i universes.
\newblock {\em Class. Quant. Grav.}, 9:S183--S185, 1992.

\bibitem{Dimastrogiovanni:2008ua}
E.~Dimastrogiovanni, W.~Fischler, and S.~Paban.
\newblock {Perturbation Growth in Anisotropic Cosmologies}.
\newblock {\em arXiv:0803.2490 [hep-th]}, 2008.

\bibitem{Gumrukcuoglu:2007bx}
A.~E. Gumrukcuoglu, Carlo~R. Contaldi, and Marco Peloso.
\newblock Inflationary perturbations in anisotropic backgrounds and their
  imprint on the cmb.
\newblock {\em arXiv:0707.4179 [astro-ph]}, 2007.

\bibitem{Pereira:2007yy}
Thiago~S. Pereira, Cyril Pitrou, and Jean-Philippe Uzan.
\newblock Theory of cosmological perturbations in an anisotropic universe.
\newblock {\em arXiv:0707.0736 [astro-ph]}, 2007.

\bibitem{Battye:2007aa}
Richard~A. Battye and Adam Moss.
\newblock Cosmological perturbations in elastic dark energy models.
\newblock {\em Phys. Rev.}, D76:023005, 2007.

\bibitem{Dunsby:1993fg}
Peter K.~S. Dunsby.
\newblock Gauge invariant perturbations of anisotropic cosmological models.
\newblock {\em Phys. Rev.}, D48:3562--3576, 1993.

\bibitem{Dunsby:1991vv}
Peter K.~S. Dunsby.
\newblock Gauge invariant perturbations in multicomponent fluid cosmologies.
\newblock {\em Class. Quant. Grav.}, 8:1785--1806, 1991.

\bibitem{Tsagas:1999ft}
Christos Tsagas and Roy Maartens.
\newblock {Magnetized cosmological perturbations}.
\newblock {\em Phys. Rev.}, D61:083519, 2000.

\bibitem{Tsagas:1999tu}
Christos~G. Tsagas and Roy Maartens.
\newblock {Cosmological perturbations on a magnetised Bianchi I background}.
\newblock {\em Class. Quant. Grav.}, 17:2215--2242, 2000.

\bibitem{Tsagas:2007yx}
Christos~G. Tsagas, Anthony Challinor, and Roy Maartens.
\newblock Relativistic cosmology and large-scale structure.
\newblock {\em arXiv:0705.4397 [astro-ph]}, 2007.

\bibitem{Ellis:1990gi}
G.~F.~R. Ellis, M.~Bruni, and J.~Hwang.
\newblock Density gradient - vorticity relation in perfect fluid
  robertson-walker perturbations.
\newblock {\em Phys. Rev.}, D42:1035--1046, 1990.

\bibitem{Koivisto:2007sq}
Tomi Koivisto.
\newblock Viable palatini-f(r) cosmologies with generalized dark matter.
\newblock {\em Phys. Rev.}, D76:043527, 2007.

\bibitem{Jimenez:2008au}
Jose~Beltran Jimenez and Antonio~L. Maroto.
\newblock A cosmic vector for dark energy.
\newblock {\em arXiv:0801.1486 [astro-ph]}, 2008.

\end{thebibliography}

\end{document}